\documentclass[journal,12pt,onecolumn]{IEEEtran}
\usepackage{amsmath}
\usepackage{amssymb}
\usepackage{graphicx}
\graphicspath{{Fig/}}
\usepackage{epstopdf}
\usepackage{caption}
\usepackage{tabu}
\usepackage{booktabs}
\usepackage{multirow}
\usepackage{multicol}
\usepackage{array}
\usepackage{bm}
\usepackage{bbm}
\usepackage{subfigure}
\usepackage{cases}
\usepackage{stfloats}
\usepackage{cite}
\usepackage{textcomp}
\usepackage{setspace}
\usepackage{color}
\begin{document}
	
\title{MDC Enhanced IoT Networks: Network Modeling and Performance Analysis}
\author{ Hongguang~Sun,~\IEEEmembership{Member,~IEEE,}~Yajun~Ma,
    Tony~Q.~S.~Quek,~\IEEEmembership{Fellow,~IEEE}\\	
	Xijun~Wang,~\IEEEmembership{Member,~IEEE,}~Kun~Guo, and~Hongming~Zhang
%
%
}

\maketitle
\vspace{-1.6cm}
\begin{abstract}
     As a promising architecture, Mobile Data Collector (MDC) enhanced Internet of Things (IoT) exhibits broad prospects in efficient data collection and data aggregation especially for sparse deployment scenarios. Combining the tools from queueing theory and stochastic geometry, we propose an analytical framework to study the network performance of an MDC enhanced IoT network, in terms of coverage probability, end-to-end delay and energy consumption. We derive the closed-form expressions for average contact and inter-contact time between a sensor and its associated MDC. By modeling the data collection system between a sensor and its associated MDCs as an M/G/1 queue system with vacations and general limited (G-limited) service, we first derive the queueing delay at the tagged sensor, and further obtain the end-to-end delay. The proposed analytical framework enables us to quantify the effect on network performance of key system parameters, such as MDC velocity, packet arrival rate, densities of sensors and MDCs, and contact radius. This study reveals that the MDC velocity has little impact on the coverage probability, and provides guidelines to minimize the end-to-end delay by optimizing the density and contact radius of sensors, and the velocity and density of MDCs.
 \end{abstract}


\section{Introduction}
\IEEEPARstart{C}{ommitted} to establishing ubiquitous connections, the Internet of Things (IoT) is in a new stage of cross-border integration, integrated innovation and large-scale deployment. A variety of upper-level applications are constructed on the basis of data generated by sensors and health monitors. For large-scale and sparsely deployed IoT scenarios, the traditional sensor network architecture can no longer meet the ever-increasing demands for data transmission due to the following issues. Firstly, sensors which are usually battery-limited consume a large amount of energy in transmitting data to distant nodes due to path loss and large-scale fading. Secondly, in traditional approach, sensors not only transmit data generated by themselves, but also forward data from neighboring nodes. Thirdly, the energy of nodes near the Access Points (APs) is exhausted much more quickly, resulting in unbalanced network energy consumption and shorter network lifetime. Therefore, it is particularly important to provide an energy-efficient and cost-efficient solution to collect data effectively. Some prior works reduced the distance between the sensor and the APs by deploying aggregators which can be considered as static relay nodes to deliver packets from neighboring sensors \cite{8768074}. However, with the continuous expansion of deployment scope in IoT networks, substantial relays should be deployed, leading to a high deployment cost and the lack of network flexibility, which motivates designers to explore new network architectures.

 In recent years, employing mobile platforms is considered as an effective approach to not only solve the problem of data collection in spatially separated areas but also maximize the lifetime of IoT networks. As a data transmission medium between sensors and APs, mobile data collectors (MDCs) \footnote{We use the term “MDCs” to differentiate from the traditional static data aggregators to highlight the moving characteristic of MDCs.} play an increasingly important role in balancing network energy consumption, especially in the scene of large-scale sensor data collection. A typical application scenario is the farmland environment monitoring of an agricultural IoT network. To provide favorable growing conditions for crops, we need to have a good knowledge of the soil conditions, such as the soil humidity, the content of soil trace elements, etc. The MDC enhanced IoT network architecture is expected to tackle the dilemma between the vastness of farmland and the high cost of certain sensors. Ubiquitous moving objects can be used to participate in network data transmission. A small number of mobile devices, such as Unmanned Aerial Vehicles (UAVs), are exclusively used for network communication by controlling their movement trajectory \cite{8859647,zhang2019stochastic,8579209}. Interestingly, in many applications, the MDCs are naturally available in the sensing field. For instance, in smart city application, a large number of vehicles dexterously play MDCs and participate in data communication \cite{anjomshoaa2018city,huang2020joint,ren2020intelligent}. In environment monitoring scenario, tourists or animals which can be equipped with micro transponders serve as MDCs \cite{sharma2020integrated,tseng2013opportunistic}. These mobile entities are non-functional for network data transmission, and thus, their movement trajectory can generally be described by a random mobility model. It is worth noting that the MDC is in charge of data collection from sensors, while the AP is responsible for the data aggregation from multiple MDCs. In practical network deployment, MDCs can be of the same type or heterogeneous, where the heterogeneity can be reflected in storage, computing capability, moving speed, etc.

\subsection{Related Work}

Most prior works on the MDC-enhanced IoT network focus on routing planning, data transmission, and storage strategy design. The authors in \cite{Ang2018OptimizingEC} employed MDCs in a large-scale sensor network, and optimized the cluster number aiming to minimize the energy consumption of sensors.
In \cite{qu2014toward}, the authors performed extensive simulations to verify the energy conservation mechanism based on the assistance of MDCs, which revealed that $86\%$ of sensors profit from 14 MDCs, and the sleep time of about $56\%$ sensors increase to more than $50\%$.
The authors in \cite{cayirpunar2017optimal} investigated the characteristics of three representative mobility patterns of MDCs, and developed a mixed integer programming framework to maximize network lifetime. In \cite{singh2018energy}, the authors proposed an MDC-assisted data collection strategy based on clusters formed by unequal and fixed grids, and optimized the time for cluster head alteration. The authors in \cite{fu2019wsns} proposed a routing mechanism and a storage management scheme for a Wireless Sensor Network (WSN)-assisted opportunistic network, with the aim to decrease the latency of message forwarding. However, all the literature mentioned above evaluates the benefit of adding MDCs only through simulations.

In practical large-scale network deployment, it is difficult to thoroughly assess the impact of all network parameters through time-consuming simulations. A rigorous theoretical analysis framework is indispensable for an MDC-assisted IoT network.
The authors in \cite{petrov2017vehicle} presented an enhanced narrowband IoT (NB-IoT) network architecture where vehicles participate in forwarding sensor traffic to the base station. In \cite{vishnuvarthan2019energy}, the authors focused on a strip-based WSN and proposed an analytical approach to analyze the average energy consumption of a sensor.
However, the works mentioned above assumed an ideal channel model, where the effects of channel fading and aggregated interference are ignored. In addition, the complete delay analysis framework in a large-scale network does not exist. The activity of sensors and MDCs, along with the random mobility of MDCs, make it difficult to characterize the distribution of interference and delay performance, especially in an opportunistic IoT networks.

In the past few years, stochastic geometry \cite{chiu2013stochastic} has been applied extensively to characterize the distribution of interference in large-scale networks, where the spatial locations of transmitters are usually modeled as classical Poisson point process (PPP). However, most of the previous works either employ the high mobility model \cite{Zhang2016DelayAR} in which the locations of users are modeled as an independent PPP in each timeslot, or the static network model \cite{8586942,qi2020traffic} in which MDCs stay relatively static with the user, while the analytical framework on the basis of a more general mobility model does not exist. In our previous conference paper \cite{miajun}, the MDCs employed a random mobility model in both data collection stage and data aggregation stage, based on which we analyzed the coverage probability. The main difference between this work and \cite{miajun} lies in the following aspects. Firstly, in \cite{miajun} we adopted a fully-loaded network model, where sensors and MDCs are assumed to always have packets to transmit. While in this work, we propose a spatiotemporal analytical model which jointly exploits the tools from queueing theory and stochastic geometry to characterize both traffic dynamics and nodes spatial randomness. Secondly, we enhance the mobility model of MDCs in data aggregation stage, which significantly increases the data transmission efficiency. Thirdly, besides coverage probability, we explore two other key network metrics in this work: end-to-end delay and energy consumption, and provide more insight for network design.


\subsection{Contributions and Organization}
     Inspired by the stated above, in this study, we propose a three-layer IoT network architecture consisting of sensors, MDCs and APs, where MDCs are served as mobile relays to assist in forwarding data from sensors to APs. A sleeping strategy is considered for sensors to save energy, where a sensor is in activity only when MDCs move into its contact area. The main contributions are listed in the following:

    \begin{itemize}
      \item By combining tools from queueing theory and stochastic geometry, we model both the spatial randomness of nodes (i.e., sensors, MDCs, and APs) and the temporal randomness of traffic, and propose a theoretical framework to analyze the network performance of large-scale IoT networks in terms of coverage probability, end-to-end delay, and energy consumption.
      \item To improve the data transmission efficiency, we propose a hybrid mobility model for MDCs, where a simple random waypoint (SRWP) mobility model is employed in the data collection stage, and a straight-line mobility model is utilized in the data aggregation stage. Under the SRWP mobility model. The closed-form of average contact time and inter-contact time between a sensor and MDCs are derived.
      \item We characterize the network interference distribution in both data collection stage and data aggregation stage, by taking into account the densities of nodes, the contact of sensors with MDCs, and the traffic arrival rate. We first derive the contact probability of the tagged sensor with MDCs, and the non-empty probability of the sensor queue, and further derive the coverage probability of the typical MDC and AP, respectively.
      \item We model the data collection system between a sensor and MDCs as an M/G/1 queue system with vacations and general limited (G-limited) service, and derive the queueing delay of the tagged packet at the tagged sensor. By further obtaining the transmission delay at the tagged sensor, the queueing delay and transmission delay at the typical MDC, we derive the end-to-end delay.
      \item The proposed analytical framework can be used to quantify the impact on network performance of key system parameters, such as velocity of MDC, packet arrival rate, densities of sensors and MDCs, and contact radius. Our results reveal that the velocity of MDC has little impact on the coverage probability, and the end-to-end delay can be minimized by optimally setting the density and contact radius of sensors, and the velocity and density of MDCs.
    \end{itemize}

    The rest of the paper is organized as follows. In section \uppercase\expandafter{\romannumeral2}, we describe the system model. Section \uppercase\expandafter{\romannumeral3} details the contact and vacation queueing process. In Section \uppercase\expandafter{\romannumeral4}, we investigate the system performance, and the accuracy of analytical model is validated with simulation. Section \uppercase\expandafter{\romannumeral5} presents the numerical results, and the effect of various network parameters on system performance are discussed. Finally, Section \uppercase\expandafter{\romannumeral6} summarizes this paper. The notations are listed in Table \uppercase\expandafter{\romannumeral1}.
    \begin{table*}
    	\caption{SUMMARY OF NOTATION}
    	\begin{center}
    		\renewcommand{\arraystretch}{1.4}
    		\begin{tabu} to 1\textwidth{|X[1,c]|X[5]|} 	
    			\hline
    			\textbf{Notation} 		&\textbf{Definition}     \\  \hline
                $\Phi_s, \Phi_m, \Phi_a$   &  Locations of sensors, MDCs, and APs modeled by three independent PPPs     \\ \hline
                $\lambda_s, \lambda_m, \lambda_a $ &Density of sensors, MDCs, and access points \\ \hline
    			$P_s, P_m$  &  Transmit powers of sensors and MDCs \\ \hline
                $\alpha,\sigma^2$   & Path loss exponent and thermal noise power  \\  \hline
                $h_x, h_y$  & Rayleigh fading channel gain from interfering sensors and from interfering MDCs \\ \hline
                $R_s, R_a$  &  Contact radius of sensors and aggregation area radius of access points  \\ \hline
    			$v, w, p$   &  Velocity, walk duration and pause duration of MDCs \\ \hline
                $T_s, T_a$  & SINR decoding threshold of MDCs and access points \\ \hline
                $K$         & Packet collection  threshold for MDCs transmitting packets to APs \\ \hline
                $\xi,\mu,\rho$       & Packet arrival rate, service rate of sensors and traffic intensity of queueing system \\ \hline
                $\delta,b_s$    & Length of a timeslot and mean service time of a packet \\ \hline
                $\lambda_m^{'}, \lambda_s^{'}$ & Spatial densities of active MDCs and sensors \\ \hline
    		\end{tabu}
    	\end{center}
    \end{table*}

\section{SYSTEM MODEL}
\subsection{Network Model}

\begin{figure}[t]
    	\subfigure[]{
    		\begin{minipage}[t]{0.5\linewidth}
    			\centering
    			\includegraphics[width=3in]{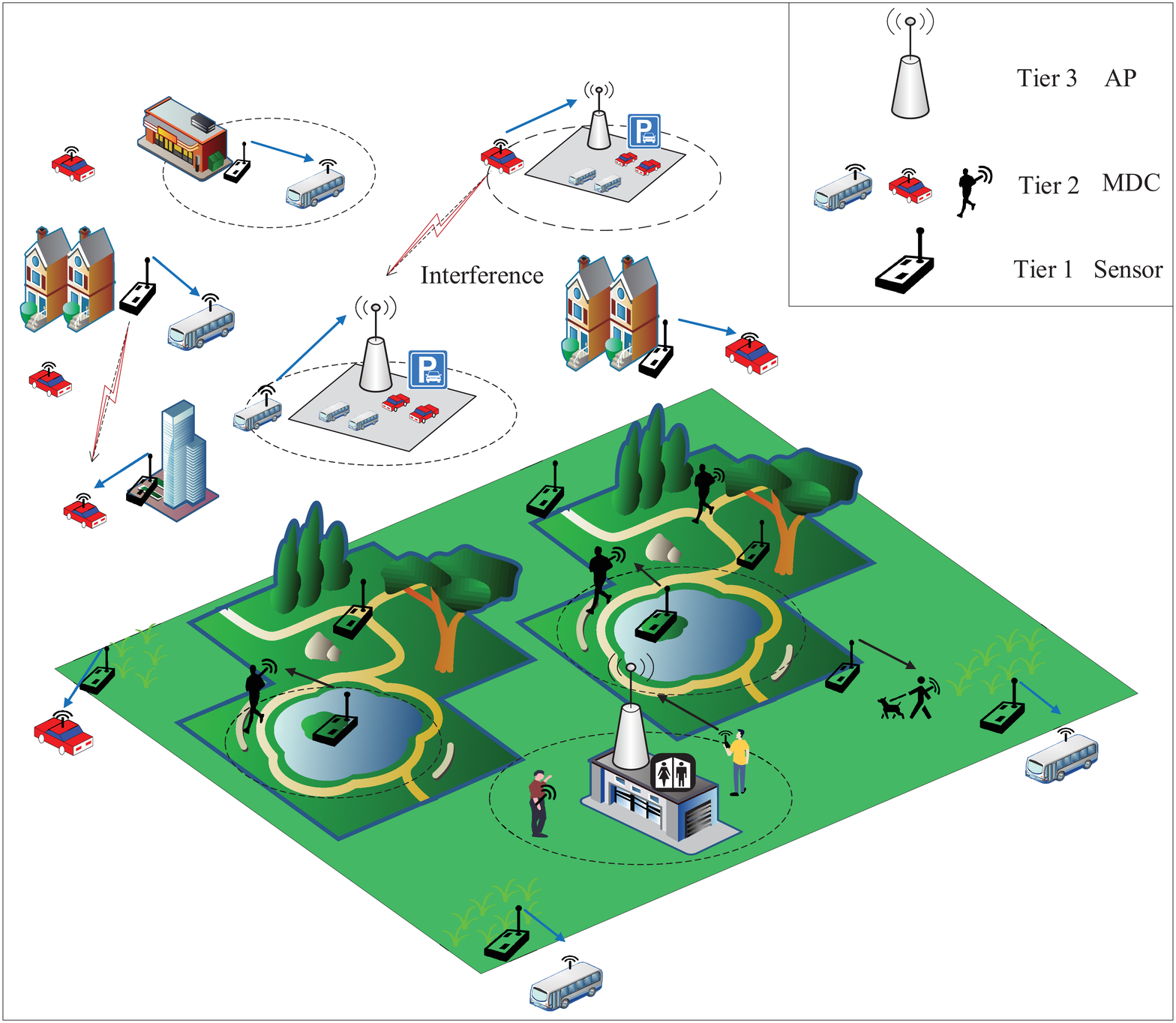}
    		\end{minipage}%
    	}
    	\subfigure[]{
    		\begin{minipage}[t]{0.5\linewidth}
    			\centering
    			\includegraphics[width = 3in]{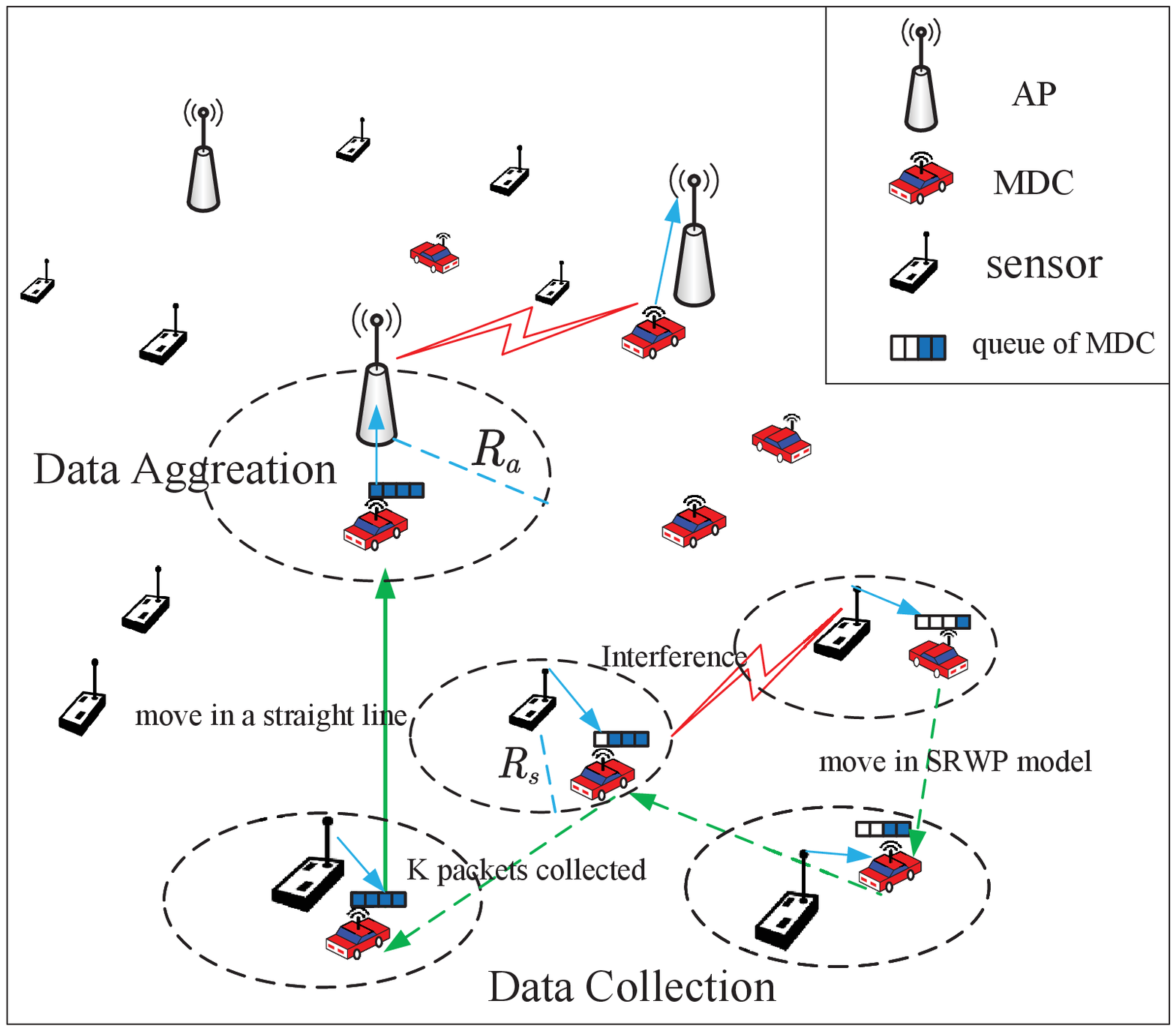}
    		\end{minipage}%
    	}
    	\caption{MDCs enhanced three-layer IoT network model}
        \label{NetworkMedel}
\end{figure}

As is shown in Fig. \ref{NetworkMedel}, the IoT network is modeled by a three-tier architecture, including APs, MDCs, and sensors, the spatial locations of which are, respectively, modeled by three independent homogeneous PPPs, defined as $\Phi_a$, $\Phi_m$, $\Phi_s$, with respective densities of $\lambda_a$, $\lambda_m$, $\lambda_s$. It is worth noting that the responsibility of MDCs is to forward packets collected by sensors to APs. \footnote{We adopt the same type of MDCs in terms of storage, computing capability and moving speed to facilitate the analysis. The heterogeneous MDCs will be considered in our future work. What’s more, we assume that the Doppler frequency shift caused by the velocity of MDCs can be eliminated using the existing techniques.}

To save the energy of sensors, we assume that the sleeping capability is equipped at a sensor who keeps sleeping unless an MDC arrives at the contact area of this sensor. The contact area is expressed as a circle with $R_s$ being the radius and the MDC is in center. When activated by an MDC, the sensor will transmit packets (backlogged in the buffer) to the MDC until the MDC roams out of the contact area. Then the sensor enters into sleeping mode again, and waits to be activated by the next coming MDC. Furthermore, to improve the reliability of data transmission, we assume a data aggregation area centered at each AP with radius $R_a$. As a certain amount of data, denoted by $K$, has been collected by an MDC, the MDC will associate with the nearest AP and moves ahead to the AP in a straight line. It finally stops at a random position within the corresponding data aggregation area and starts to forward the data to the AP. The MDCs are assumed to be equipped with a battery with enough storage so as to support the roaming, data collection and transmission.

\textbf{Remark 1.} {\itshape Compared to the Poisson cluster process, the merit of modeling the spatial locations of MDCs as a homogeneous PPP is twofold. On the one hand, it increases the contact opportunities of a sensor with MDCs, and enhances the efficiency of packet forwarding. On the other hand, the PPP assumption significantly simplifies the performance analysis, and the derived closed-form expressions provide more insights for the system design.}

\subsection{Traffic Model and Transmission Strategy}

The packet arrival at each sensor is subject to a Poisson process with average arrival rate being $\xi$. We consider the case where each sensor has a queue of adequate capacity for accommodating the arriving packets. Once an MDC enters the contact area of a sensor, the sensor begins to transmit packets to the MDC based on the First-Come-First-Served (FCFS) discipline. When a packet is successfully received by the MDC, the sensor will receive an ACK from the MDC on a separate feedback channel, and remove the packet from the buffer. Otherwise, the MDC sends a NACK, and the packet still queues up and waits to be retransmitted. The ACK/NACK transmission is assumed to be instantaneous and error-free \cite{7544562}. We define the transmission cycle as the summation of a contact duration and the consecutive inter-contact duration. Thus, the $k$-th transmission cycle begins from the k-th contact duration and ends at the $k$-th subsequent inter-contact duration. To simplify the analysis, we consider the following transmission strategy: the packets arrive at the $k$-th transmission cycle can only be transmitted in the $(k+1)$-th transmission cycle. Note that this assumption is reasonable, especially for the case when the contact duration is much smaller than the inter-contact duration. In such case, the queueing packets arrive in the last transmission cycle are unable to be cleared during the contact duration of the current transmission cycle. The time is assumed to be divided into equal-sized slots of length $\delta$. We consider the constant bit rate coding, and assume that a packet can be transmitted exactly within a timeslot.
Because of the random channel fading and existing interference, the successful reception of a packet may need multiple transmissions.

\subsection{Mobility Model}

In this work, we consider the following mobility model. In the data collection stage, an MDC follows the SRWP mobility model presented in \cite{Banagar2019FundamentalsOD}. Once the MDC has collected $K$ packets, it will enter the data aggregation stage where the MDC moves ahead to its associated AP in a straight line, and stop within the data aggregation area to forward packets to the AP. We first give the definition of SRWP model in the following.

\textbf{Definition 1.} (SRWP). At the start, each MDC keeps still at the initial position for a fixed period $p$. Then, It independently chooses a direction according to the uniform distribution $\theta \sim U[0,2\pi]$, and moves along the direction for a fixed period $w$ at a constant velocity $v$. When stoping at a new position, the MDC keeps still for another period $p$ before randomly choosing another direction and repeating the above procedure.

As a special case of SRWP, a straight line mobility model without stopping and changing direction \cite{3gpp2018} has been employed by 3GPP to characterize the mobility of drones. With the proposed SRWP mobility model, we can derive the probability density function (PDF) of the distance $r_0$ from a sensor to its connected MDC as
\begin{equation}\label{FML_fr0}
      \begin{aligned}
    f_{r}\left(r_{0}\right)=\left\{\begin{array}{ll}
    \frac{2 r_{0}}{R_{\mathrm{s}}^{2}}, &r_{0} \in\left[0, R_{\mathrm{s}}\right] \\
    	\ 0,& \text{ otherwise }
    \end{array}\right..
    \end{aligned}
\end{equation}

\subsection{Channel modelling and Interference characterization}

For the channel model, we consider both large-scale path loss and small-scale fading. According to Slivnyak's theorem \cite{chiu2013stochastic}, we can employ the performance of a typical receiver to represent the average network performance. Without loss of generality, we assume that the typical receiver is located at the origin, and the corresponding transmitter is located at a distance $r_0$ away. We can derive the received power at the typical receiver as $P h_0r_0^{-\alpha}$, where $P$ represents transmit power and $h_0$ denotes small-scale fading power gain. In this work, we consider Rayleigh fading, and thus, $h_0$ is subject to the exponential distribution with unit mean, i.e., $h_0 \sim \rm exp(1)$. In addition, we define $\sigma^2$ as the variance of the additive white Gaussian noise.

In this work, MDCs and sensors are assumed to operate on the orthogonal frequency bands. As such, during the data collection (aggregation) stage, the typical MDC (AP) suffers from interference originated from active sensors (MDCs), which can be expressed as
\begin{equation}\label{FML_interf_MDC}
  I_{r}^{m}=\sum_{x\in\Phi_{s}\backslash\left\{s_{0}\right\}} \mathbbm{1}_x P_{s} h r_{x}^{-\alpha},
\end{equation}

\begin{equation}\label{FML_interf_APs}
  I_{r}^{a}=\sum_{y \in \Phi_{b}}\mathbbm{1}_y P_{m} h r_{y}^{-\alpha},
\end{equation}
where $\mathbbm{1}_x $ in Eq. (\ref{FML_interf_MDC}) is the indicator function where $\mathbbm{1}_x = 1$ holds if the interfering sensor located at $x$ is active, while $\mathbbm{1}_x = 0$ holds otherwise. The symbol $h$ denotes the channel power gain. In addition, $r_x$ ($r_y$) denotes the distance between the interfering sensor (MDC) to the typical MDC (AP).

\section{CONTACT AND QUEUEING MODELING }

In this section, we will characterize the contact process between the MDC and sensors, and the packets queueing process in the system. The results obtained in this section will be applied to the system performance analysis in the next section.

\subsection{Contact Model Characterization}

To characterize the contact distribution between sensors and MDCs, the timeline is divided into contact time and inter-contact time. The time that MDCs sojourn in the contact area of sensors is defined as contact time, while the time period between two adjacent contact durations is defined as inter-contact time. The discrete contact events between an MDC and sensors can be modeled by an alternating renewal process \cite{Ziegel2004SystemRT} as follows.

\textbf{Definition 2.} Let $\{X(t), t\geq 0\}$ a stochastic process with only $0$ and $1$ in the state space, where state $1$ and state $0$ denote the discrete events of contact and inter-contact, respectively. We use $CT_k$ and $ICT_k$ ($k = 1, 2, \cdots$) to represent the $k$-th successive contact period and inter-contact period, respectively. Define $\psi_k = CT_k +ICT_k$, and $\psi_k$ can be referred to as an alternating renewal process.

Define ${\rm E}(ICT)$ and ${\rm E}(CT)$ as the average inter-contact time and contact time, which are dependent on various factors, e.g., MDCs' velocity, a sensor's contact area, the walk and pause duration of the MDC, etc. For a typical sensor, we define $\mathbb{P}_{ct}$ as the contact probability with an MDC. According to the theory of alternate renewal process, each time the system transitions from state $0$ to state $1$ is a ``regeneration point" of the process. The process develops from that moment onward as if the process started from the beginning (without being affected by the history of the process prior to this moment). Thus, we have

\begin{equation}\label{FML_active1}
  \mathbb{P}_{ct} =\lim_{t\rightarrow \infty}Pr\left\{ X\left( t \right) =1 \right\} = \frac{{\rm E}(CT_k)}{{\rm E}(CT_k)+{\rm E}(ICT_k)} = \frac{{\rm E}(CT)}{{\rm E}(CT)+{\rm E}(ICT)},
\end{equation}
where ${\rm E}(CT)$ and ${\rm E}(ICT)$ are given by Lemma 1 and Lemma 2, respectively.

\textbf{Lemma 1.} {\itshape  The expectation of contact time ${\rm E}(CT)$ between a sensor and MDCs under the SRWP model can be obtained by}
\begin{equation}\label{FML_ECT}
  {\rm E}(CT) = \frac{\pi R_s \left[2v(w+p)+4{\rm E}\left(D\right)-\pi R_s\right]}{4wv^{2}},
\end{equation}
{\itshape where p and w, respectively, denote the pause duration and walk duration, and ${\rm E}(D)$ is given by}
\begin{equation}\label{FML_ED}
 {\rm E}\left(D\right)=\frac{1}{2 \pi^{2} R_{s}^{2}} \int_{0}^{2 \pi} \int_{0}^{2 \pi} \int_{0}^{R_{s}}
  \sqrt{\left(r \cos \alpha-R_{s} \cos \theta\right)^{2}+\left(r \sin \alpha-R_{s} \sin \theta\right)^{2}} r dr d\alpha d\theta.
\end{equation}

{\itshape Proof:} See Appendix A.

\textbf{Lemma 2.} {\itshape The expectation of inter-contact time ${\rm E}(ICT)$ between MDCs and a sensor under the SRWP model can be obtained by}
\begin{equation}\label{FML_EICT}
  {\rm E}(ICT) = \frac{1}{\epsilon} = \frac{w+p}{2 w v \lambda_m  R_s},
\end{equation}
{\itshape where $\epsilon$, $\lambda_m$, and $v$, respectively, represent the arrival rate, density and velocity of MDCs, and $R_s$ denotes the radius of the sensor's contact area.}

{\itshape Proof}. The result in (\ref{FML_EICT}) is derived by the modifying Theorem 3.4 in \cite{spyropoulos2006performance}, where the length of each epoch $L = wv$, and the expected duration of each epoch is equal to $w+p$.


\begin{figure}[t]
    	\subfigure[]{
    		\begin{minipage}[t]{0.5\linewidth}
    			\centering
    			\includegraphics[width=2.8in]{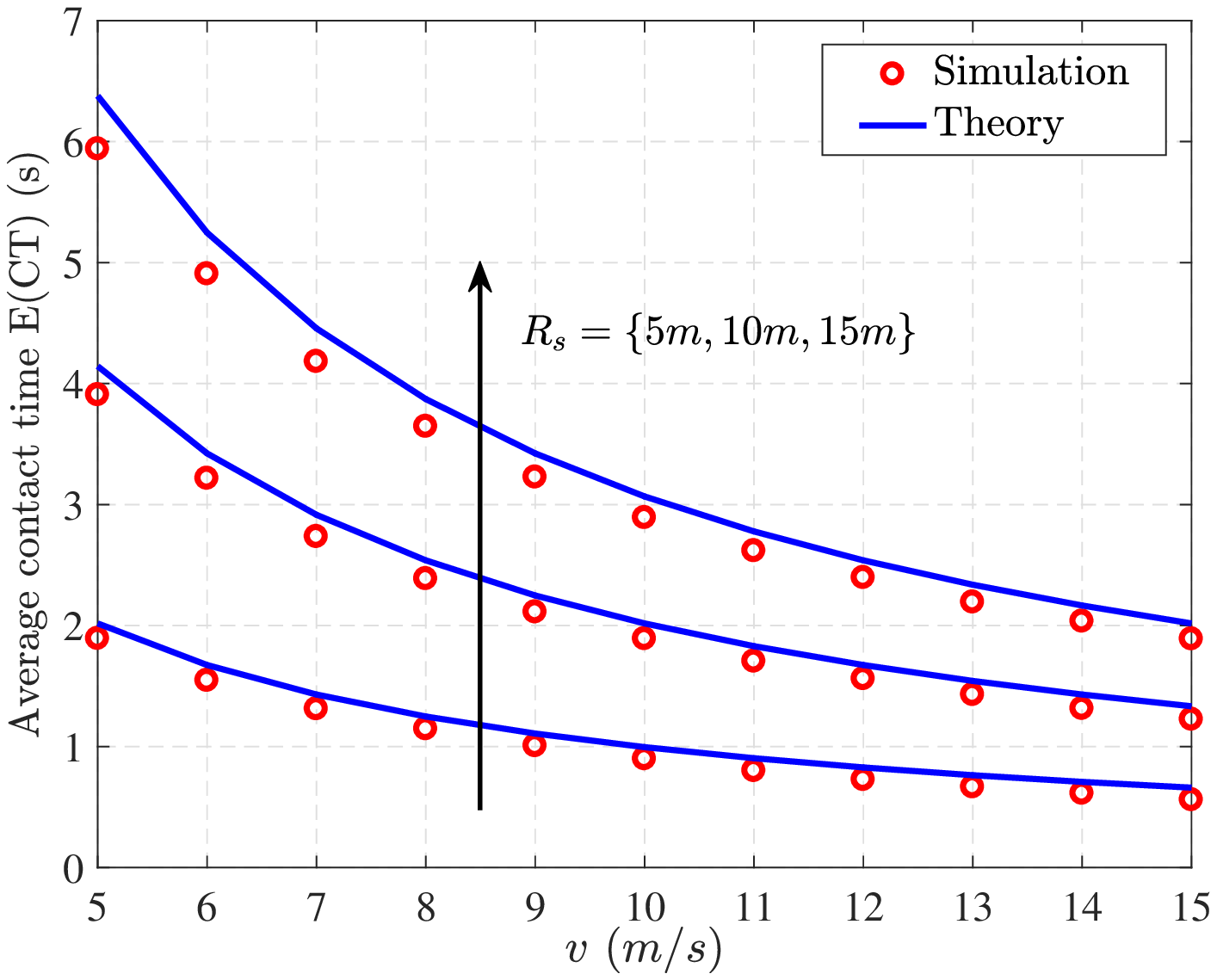}
    		\end{minipage}%
    	}
    	\subfigure[]{
    		\begin{minipage}[t]{0.5\linewidth}
    			\centering
    			\includegraphics[width = 2.8in]{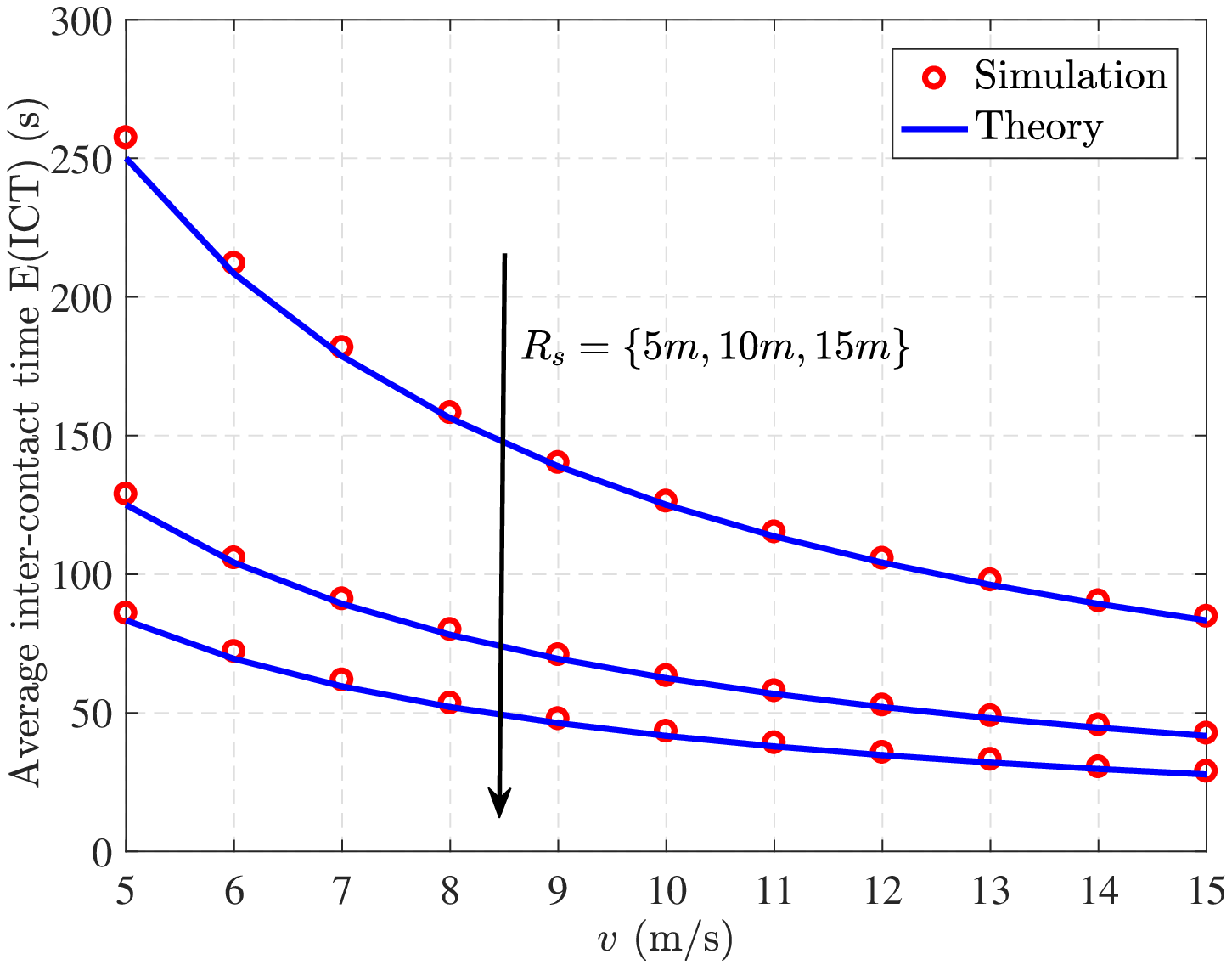}
    		\end{minipage}%
    	}
    	\caption{Comparison of simulations and analytical results for (a) ${\rm E}(CT)$ and (b) ${\rm E}(ICT)$ v.s. MDC's  velocity $v$ for different $R_s$, with $\lambda_s = 10^{-3}m^{-2}$, $\lambda_m = 10^{-4}m^{-2}$.}
        \label{aveCT_ICT}
\end{figure}
\begin{figure}
	\centering
	\includegraphics[width=2.8in]{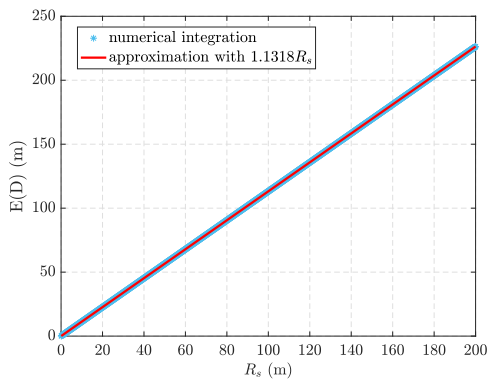} 
	\caption{Numerical integration and approximation of ${\rm E}(D)$ v.s. contact radius $R_s$.} 
	\label{EDfit} 
\end{figure}
We show the accuracy of ${\rm E}(CT)$ and ${\rm E}(ICT)$, respectively, given in Eq. (\ref{FML_ECT}) and Eq. (\ref{FML_EICT}) in Fig. \ref{aveCT_ICT}, by comparing with simulation results. We observe that both ${\rm E}(CT)$ and ${\rm E}(ICT)$ decrease as MDC's velocity grows up. This is due to the fact that a growing velocity not only decreases the MDC's sojourn time within the contact area, but also reduces the time within the inter-contact area. In addition, a larger contact radius $R_s$ leads to an increase in ${\rm E}(CT)$ and a decrease in ${\rm E}(ICT)$. This is due to the growing sojourn time of MDCs within a larger contact area.

Substituting Eq. (\ref{FML_ECT}) and Eq. (\ref{FML_EICT}) into Eq. (\ref{FML_active1}), we obtain the probability that the typical sensor is in contact with an MDC, referred to as the contact probability, as below.
\begin{equation}\label{EqCtPro}
        \mathbb{P}_{\rm ct} = \frac{2 \pi R_s^{2} v \lambda_m[2 v(w+p)+4 {\rm E}(D)-\pi R_s]}{2 \pi R_s^{2} v \lambda_m[2 v(w+p)+4 {\rm E}(D)-\pi R_s]+4 (w+p) v^{2}}
        \mathop{\approx}\limits^{(a)}\frac{1}{1 + \frac{1}{\pi R_s^{2}\lambda_m[1+\frac{0.6428R_s}{v(w+p)}]}},
\end{equation}
where step (a) is obtained by approximating ${\rm E}(D)$ to $1.1318R_s$. To simplify the expression of $\mathbb{P}_{\rm ct}$ and further reveal the impacts of key system parameters on contact probability, we approximate ${\rm E}(D)$ by using a numerical fitting approach. It is shown from Fig. \ref{EDfit} that the approximation has a high accuracy when $R_s$ varies.

\textbf{Remark 2. }{\itshape It can be seen from Lemma 1 and Lemma 2 that with the increase of epoch time $w$ and velocity of MDC $v$, both ${\rm E}(CT)$ and ${\rm E}(ICT)$ decrease. Meanwhile, with the increase of pause time $p$, both ${\rm E}(CT)$ and ${\rm E}(ICT)$ go up. Due to the fact that $v(w+p) \gg 0.6428R_s$ in the practical deployment, accoording to eq. (8), the change in  $w$, $p$ and $v$ have little impact on contact probability $\mathbb{P}_{\rm ct}$ and coverage probability.}


\subsection{Queueing Model Characterization}

\begin{figure*}
    	\centering
    	\includegraphics[width=6in]{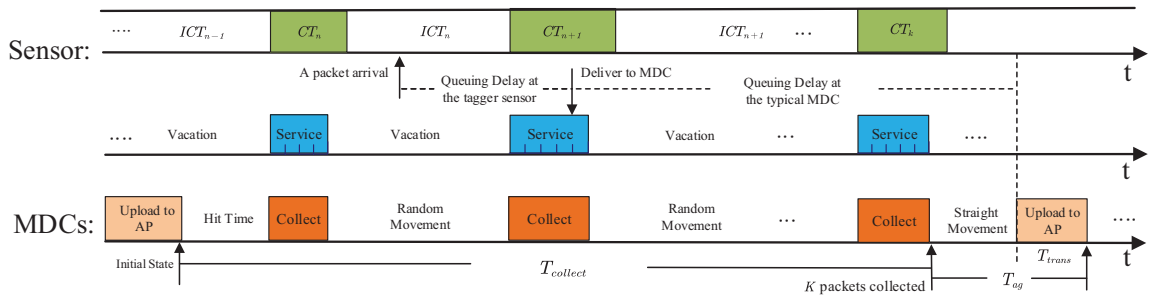} 
    	\caption{Illustration of contact model and vacation queueing model between a sensor and MDCs.} 
    	\label{vacationModel}
    \end{figure*}

    As is depicted in Definition 2, the contact process between a sensor and an MDC is an alternating renewal process. If an MDC and a sensor are in contact, packets in the queue of the sensor are uploaded to the MDC according to the FCFS discipline. If an MDC and a sensor are in the inter-contact state, we consider that the sensor is on vacation and packets in queue are waiting for service. Due to the limited contact time, the sensor will be mandatory to take a vacation after it serves a certain number of packets in a service period within a contact period. It is worth noting that the service period is smaller than the contact period due to the fact that the sensor’s queue may be cleared up at a certain moment within the contact period, as shown in Fig. \ref{vacationModel}. The data collection system between a sensor and MDCs can be modeled by an M/G/1 vacation queueing system with general limited (G-limited) service \cite{tian2006vacation}, where sensors are regarded as servers and packets are served as customers.

    The service time of each packet in the queue of the sensor is i.i.d. with the mean denoted by $b_s$. The mean service rate is given by $\mu = 1/b_s$ and the traffic intensity can be denoted by $\rho = \xi/\mu$. We focus on the queueing system of the tagged sensor, and assume that at the end of $ICT_n$, the tagged sensor serves $\Psi$ packets during $CT_{n+1}$. We assume that $\Xi$ is the upper bound of the number of packets collected by MDCs during a service period, which is determined by both average contact time ${\rm E}(CT)$ and mean service rate $\mu$. When the average service time equals to the average contact time ${\rm E}(CT)$, the average number of packets collected by MDCs during each contact period reaches the maximum, thus we have
    \begin{equation}\label{Eq_Xi}
      \Xi = \lfloor \mu {\rm E}(CT) \rfloor.
    \end{equation}
     According to transmission strategy described in Section B of the system model, the packets arrive at the k-th transmission cycle can only be transmitted in the (k + 1)-th transmission cycle. Therefore, when the queueing system reaches steady state, the average number of packets transmitted within the contact time can be given by
    \begin{equation}\label{E_Psi}
      {\rm E}(\Psi) = \xi[{\rm E}(CT) + {\rm E}(ICT)].
    \end{equation}
    Thus, the average service time can be expressed as
    \begin{equation}
        {\rm E}(\mathcal{S}) = \rho[{\rm E}(CT) + {\rm E}(ICT)],
    \end{equation}
    and the average vacation duration can be expressed as
    \begin{equation}\label{ave_vacation}
      {\rm E}(V) = (1-\rho)[{\rm E}(CT) + {\rm E}(ICT)].
    \end{equation}

     \textbf{Remark 3.} {\itshape A queueing system can reach steady state if and only if the packet arrival rate $\xi$ is less than the service rate $\mu$. For a vacation queueing system with G-limited service, the queue can reach steady state if and only if the average number of packets arriving during a transmission cycle, including a contact duration and the consecutive inter-contact duration, are less than the average number of packets served during the contact period. Therefore, the stability condition of the queueing system is given by
       \begin{equation}\label{stable_eq}
            \rho = \frac{\xi}{\mu}  < \frac{{\rm E}(CT)}{{\rm E}(CT) + {\rm E}(ICT)} = \mathbb{P}_{\rm ct}.
        \end{equation}
    }
    From Eq. (\ref{stable_eq}), we know that if the traffic intensity $\rho$ is less than the contact probability $\mathbb{P}_{\rm ct}$, the queueing system will remain stable. Transforming Eq. (\ref{stable_eq}), we can obtain the upper bound on the arrival rate of sensor packets in the steady state of the system, which is given by
    \begin{equation}
            \xi  < \frac{\mu {\rm E}(CT)}{{\rm E}(CT) + {\rm E}(ICT)}.
     \end{equation}

     As depicted in Fig. \ref{vacationModel}, the vacation duration is mainly composed of the inter-contact time which follows an exponential distribution with parameter ${\rm E}(ICT)$ given in Lemma 2, thus, we can approximate the vacation distribution of the tagged sensor with an exponential distribution with a mean ${\rm E}(V)$. By substituting Eq. (\ref{FML_ECT}) and Eq. (\ref{FML_EICT}) into (\ref{E_Psi}) and (\ref{ave_vacation}), we can derive ${\rm E}(\Psi)$ and ${\rm E}(V)$, respectively.

    We assume that at the end of $ICT_n$, the queue length in the sensor is $L_{n}^{\ast}$, and the steady-state distribution of $\{L_n^{\ast},n = 1,2,3,\dots\}$ is denoted by $q_k$.

    \textbf{Lemma 3.} {\itshape When the vacation queueing system with G-limited service is in steady state and the vacation time follows an exponential distribution, the average queue length is given by }
    \begin{equation}\label{Eq_EL}
       {\rm E}(L) = {\rm E}(L_{M/G/1}) + {\rm E}(L^\ast) = \rho + \frac{\xi^2 b^{(2)}}{2(1-\rho)} + {\rm E}(L^\ast).
    \end{equation}
    {\itshape where ${\rm E}(L_{M/G/1})$ represents the average queue length of the classical M/G/1 queueing model, and $b^{(2)}$ is the second moment of the service time (in timeslots) for a packet. The ${\rm E}(L^\ast)$ in (\ref{Eq_EL}) is the average number of packets at the start of each service period, which is given by}
    \begin{equation}\label{Eq_ELast}
      \begin{aligned}
        {\rm E}(L^{\ast}) &=\frac{{\rm E}\left( \Psi \right) \left[ \Xi +2\left( 1-\rho \right) {\rm E}\left( \Psi \right) \right]}{\Xi -{\rm E}\left( \Psi \right)}+\frac{\xi ^2b^{\left( 2 \right)}{\rm E}\left( \Psi \right)}{2\left( 1-\rho \right) \left[ \Xi-{\rm E}\left( \Psi \right) \right]}\\
                    &\quad- \frac{\left( 1+\rho\right)\left\{ Q_{\Xi}^{\left( 2 \right)}\left( 1 \right) +\Xi \left( \Xi -1 \right) \left[ 1-Q_{\Xi}\left( 1 \right) \right] \right\}}{2\left[ \Xi - {\rm E}(\Psi)\right]}
        \end{aligned}
    \end{equation}
    {\itshape where $Q_{\Xi}\left( z \right) \triangleq \sum_{k=0}^{\Xi -1}{q_kz^k}$, and the coefficients of $Q_{\Xi}\left( z \right)$ can be derived by solving the equations which is derived by}
    \begin{numcases}{}
         \sum_{k=0}^{\Xi-1} q_k\left\{ z_{m}^{\Xi}\left[ {\rm B}^*\left( \xi -\xi z_m \right) \right] ^k - \left[ {\rm B}^*\left( \xi -\xi z_m \right) \right] ^\Xi \right. \notag \\
          \qquad\quad\ \left. \cdot z_{m}^{k} \right\} \  = 0 \\
         \sum_{k=0}^{\Xi-1}{\left( \Xi-k \right) q_k}=\Xi-\frac{\xi {\rm E}\left( V \right)}{1-\rho} \label{eq_2}
        \end{numcases}
    $$
    z_m =\sum_{n=1}^{\infty}\left\{\frac{e^{2\pi mnj/\Xi}}{n!}\cdot \frac{d^{n-1}}{dz^{n-1}}\left\{ {\rm V}^*\left( \xi -\xi z \right) \cdot\left[{\rm B}^*(\xi -\xi z) ^\Xi \right] \right\}^{\frac{n}{\Xi}}\mid_{z=0} \right\}.
    $$

    {\itshape Proof:} The main proof steps of this conclusion have been given in Section 3.3.2 of work \cite{tian2006vacation}.

    The ${\rm V}^*(z)$ denotes the Laplace-Stieltjes Transform (LST) of the vacation duration, and is given by
    \begin{equation}\label{Eq_LST_Vz}
      {\rm V}^\ast(z) = \frac{1}{z {\rm E}(V) + 1}.
    \end{equation}
    The ${\rm B}^*(z)$ denotes the LST of the service time, and it is related to the coverage probability of the MDC which will be analyzed in the next section.

\section{PERFORMANCE ANALYSIS}

In this section, we analyze the impact of mobility model and traffic model on the network performance in terms of coverage probability, end-to-end delay and energy consumption.

We define the coverage probability or, equivalently, the service rate as the probability that the SINR received by a typical receiver is larger than a predefined SINR threshold $T$, which can be expressed as
\begin{equation}\label{FML_SINR_define}
   \mathbf{P}_{\rm cov} \triangleq \mathbb{P}[SINR>T].
\end{equation}
With regards to MDCs and APs, we redefine $\mathbf{P}_{\rm cov}^{\rm M}$ and $\mathbf{P}_{\rm cov}^{\rm A}$ as the coverage probability of the typical MDC and the typical AP.

\subsection{Coverage Probability of a typical MDC}

Given that the tagged sensor is located at a distance $r_0$ away, the received SINR of the typical MDC, which is randomly selected and assumed to be located at the origin is given by
\begin{equation}\label{FML_SINR_define2}
  SINR_m = \frac{P_{s} h r_{0}^{-\alpha}}{I_{r}^m + \sigma^{2}},
\end{equation}
where the aggregated interference $I^m_r$ is the summation of the received interference originated from all the other active sensors.

As depicted in the traffic model in Section \uppercase\expandafter{\romannumeral2}, the arrival of packets at each sensor follows a Poisson process with rate $\xi$, and the mean service time for delivering a packet is given by $b_s = \frac{\delta}{\mathbf{P}_{\rm cov}^{\rm M}}.$
For a typical MDC to collect packets from a tagged sensor, the following two conditions should be satisfied: i) the typical MDC enters into the communication range of the tagged sensor, and ii) the queue of the sensor is non-empty. In the following, the non-empty probability of the queue is derived by
                \begin{equation}\label{pr_nonEmpty}
                    \mathbb{P}_{q} = \frac{{\rm E}(\mathcal{S})}{{\rm E}(CT)} =\frac{min\left\{b_s{\rm E}(\Psi),{\rm E}(CT)\right\}}{{\rm E}(CT)}
                    = \frac{min\left\{{\delta \xi[{\rm E}(CT) + {\rm E}(ICT)] }/{\mathbf{P}_{\rm cov}^{\rm M}},{\rm E}(CT)\right\}}{{\rm E}(CT)}.
                \end{equation}
Furthermore, we can obtain the active probability of sensors
                 \begin{equation}\label{eq_pr_act_s}
                    \mathbb{P}_{act}^{s} = \mathbb{P}_{ct}\cdot\mathbb{P}_{q} = \frac{min\left\{{\delta \xi[{\rm E}(CT) + {\rm E}(ICT)] }/{\mathbf{P}_{\rm cov}^{\rm M}},{\rm E}(CT)\right\}}{{\rm E}(ICT) + {\rm E}(CT)},
                 \end{equation}
where $\mathbb{P}_{ct}$ is the contact probability of a tagged sensor, which is given in (\ref{EqCtPro}).

With the derived active probability of sensors $\mathbb{P}_{act}^{s}$,  we can determine the density of interfering sensors, which will be used in the derivation of coverage probability.

In the following, we derive the coverage probability of a typical MDC in Theorem 1.

\textbf{Theorem 1.} {\itshape The coverage probability of a typical MDC in the communication range of the tagged sensor is:}\\
\begin{equation}\label{FML_cov1}
   \mathbf{P}_{\rm cov}^{ \rm M}=
   \int_0^{R_s}{\!}\!\exp \left( -\frac{T_s r_{0}^{\alpha}\sigma ^2}{P_s}\!\!-2\pi \!\int_{r_0}^{\infty}{\frac{\lambda_{s}^{'}T_s r_x}{T_s +\left( \frac{r_x}{r_0} \right) ^{\alpha}}}dr_x \right) \cdot \frac{2r_0}{R_{s}^{2}}dr_0
\end{equation}
{\itshape where $\lambda_{s}^{'}$ is the density of active interfering sensors given by}
\begin{equation}\label{Eq_act_lamdas}
   \lambda_{s}^{'} = \mathbb{P}_{act}^{s}\lambda_s.
\end{equation}

{\itshape Proof:} See Appendix D.

\subsection{Coverage Probability of a typical AP}

Similarly, the SINR of a typical AP can be expressed as:
\begin{equation}\label{FML_SINR_define3}
  SINR_a = \frac{P_{m} h r_{0}^{-\alpha}}{I_{r}^a + \sigma^{2}},
\end{equation}
where the aggregated interference $I_{r}^a$ is given by (\ref{FML_interf_APs}).

From an MDC's perspective, it periodically alters among the following two stages: data collection stage and data aggregation stage, the average time of which are expressed as ${\rm E}(T_{\rm collect})$ and ${\rm E}(T_{\rm ag})$, respectively, as shown in Fig. \ref{vacationModel}. In order to collect $K$ packets from sensors, an MDC may need to experience multiple contact durations (denoted by $N_c$) with different sensors, which can be derived by
\begin{equation}\label{eq_Nc}
  N_c = \frac{K}{{\rm E}(\Psi)} = \frac{K}{\xi[{\rm E}(CT)+{\rm E}(ICT)]}.
\end{equation}
Note that the spatial distribution of sensors follow a homogeneous PPP $\Phi_s$ with density $\lambda_s$, and thus, the average time interval for an MDC to contact with two consecutive sensors (denoted by ${\rm E}(ICT_{s})$) can be derived similar to Lemma 2, given by
\begin{equation}\label{eq_EICT_sensors}
    {\rm E}(ICT_{s}) = \frac{w+p}{2wv\lambda_s R_s}.
\end{equation}
The difference between ${\rm E}(ICT_{s})$ in (\ref{eq_EICT_sensors}) and ${\rm E}(ICT)$ in (\ref{FML_EICT}) lies in the fact that $\lambda_m$ is replaced by $\lambda_s$  in the denominator of (\ref{eq_EICT_sensors}). Furthermore, we can obtain
\begin{equation}\label{eq_ET_collect}
        {\rm E}(T_{\rm collect}) = {\rm E}(HT) + N_c {\rm E}(CT) +(N_c-1) {\rm E}(ICT_s)
        = \frac{K\left\{{\rm E}(CT) + {\rm E}(ICT_s)\right\}}{\xi({\rm E}(CT) + {\rm E}(ICT))} - \frac{{\rm E}(ICT_s)}{2},
\end{equation}
where ${\rm E}(HT)$ is the average time taken by an MDC to contact with the first sensor from initial state, as shown in Fig. \ref{vacationModel}, and it can be derived by
    \begin{equation}
 	{\rm E}(HT) = \frac{{\rm E}(ICT_s)}{2}.
    \end{equation}

When the MDC is in data aggregation stage, the average time it takes for an MDC to move straight to the nearest associated AP is denoted by ${\rm E}(T_{\rm Smov})$, and we can obtain
\begin{equation}\label{eq_ET_Smov}
        {\rm E}(T_{\rm Smov}) = \int_{R_a}^{\infty}{\frac{r}{v} 2\pi\lambda_b r \exp\left(-\pi \lambda_b r^2 \right)}dr.
\end{equation}
When the MDC reaches the aggregation area of the AP, the duration for the MDC transmitting packets to the typical AP is given by
\begin{equation}\label{eq_ET_Trans}
  {\rm E}(T_{\rm trans}) = \frac{K \delta}{\mathbf{P}_{\rm cov}^{\rm A}}.
\end{equation}
Therefore, the duration in the data aggregation state can be obtained by
\begin{equation}\label{eq_ET_return}
        {\rm E}(T_{\rm ag}) = {\rm E}(T_{\rm Smov}) + {\rm E}(T_{\rm trans}).
\end{equation}
Specifically, for a typical AP that are receiving packets from an MDC, the interference is caused by the other active MDCs that are transmitting packets to their associated APs. When the system reaches steady state, the active probability of an MDCs can be expressed as
\begin{equation}\label{eq_pr_MDC_trans}
  \mathbb{P}_{\rm act}^{\rm M}=\frac{{\rm E}\left( T_{\rm trans} \right)}{{\rm E}\left( T_{\rm collect} \right) + {\rm E}(T_{\rm ag})}.
\end{equation}

According to \cite{Banagar2019FundamentalsOD}, after moving with the SRWP model, the spatial position of the displaced points form another homogeneous PPP with the same density. In addition, Due to the assumption that the MDC in data aggregation stage stops at a random position in the aggregation area centered by the associate AP, and the independent and identically distributed characteristics of MDCs displacement, the MDCs in data transmission stage follows another homogeneous with the thinned density $\lambda_m^{'}$ by using the {\itshape displacement theorem} in \cite{Haenggi2012StochasticGF}. In the following theorem, we derive the coverage probability of a typical AP.

\textbf{Theorem 2.} {\itshape The coverage probability of a typical AP can be obtained by}
\begin{equation}\label{FML_cov2}
    \mathbf{P}_{\rm cov}^{\rm A} = \int_0^{R_a}{\frac{2r_0}{R_a^2}}\!\exp \left(-\frac{T_a r_{0}^{\alpha}\sigma ^2}{P_{m}}
    - 2\pi \int_0^{R_a}{\frac{2r_x}{R_a^2}}\int_{r_x}^{\infty}{\frac{\lambda_m^{'}T_a  u}{T_a + \left( \frac{u}{r_0} \right)^{\alpha}}}du dr_x \right) dr_0,
\end{equation}
{\itshape where
\begin{equation}\label{Eq_act_lamdam}
        \lambda_m^{'} = \mathcal{A}_b \mathbb{P}_{\rm act}^{\rm M}\lambda_b,
\end{equation}
 and $\mathcal{A}_b$ denotes the probability of an AP being associated which is given by}
\begin{equation}\label{Eq_Ab}
        \mathcal{A}_b = 1 - \left( 1+\frac{\lambda_m}{3.5\lambda_b}\right)^{-3.5}.
\end{equation}

{\itshape Proof}. The proof of Eq. (\ref{FML_cov2}) is similar to that of $\mathbf{P}_{\rm cov}^{\rm M}$ in Appendix B, and is omitted here. The Eq. (\ref{Eq_act_lamdam}) can be explained by the fact that, there is a one-to-one mapping between a transmitting MDC and its associate AP, and we assume that $\lambda_m > \lambda_b$. The proof of Eq. (\ref{Eq_Ab}) can be found in \cite{7073589}.

\subsection{Delay Performance}

In this part, we analyze the delay performance of the proposed IoT network with MDCs, and take the end-to-end delay of a packet as the performance metric. The end-to-end delay is defined as the average time it takes for a packet to be received by the AP from the moment it is generated by the sensor. Generally, the end-to-end delay includes the following four parts: queueing delay, transmission delay, processing delay and propagation delay. In this work, processing delay and propagation delay are negligible, and thus, we focus on the queueing delay and transmission delay in our analysis. Taking into account the queueing position of a packet, the queueing delay can be divided into two parts: i) queueing delay in sensor, and ii) queueing delay in MDC.

In the following, we observe a randomly chosen packet, referred to as the tagged packet, within the queue of the tagged sensor, and attach a counter to the tagged packet. It is worth noting that when the tagged packet arrives at the queue of the tagged sensor, the counter is initiated, and when the tagged packet is successfully received by an AP, the counter is stopped. We will sequentially analyze the delay experienced by the tagged packet in the following.

\subsubsection{Queueing Delay at the Tagged Sensor}

    This part of delay is defined as the time interval from the arrival of the tagged packet at the queue of the tagged sensor to the first transmission attempt to an MDC, as shown in Fig. \ref{vacationModel}. According to the transmission strategy in Section II.B, when the system reaches steady state, the queueing delay of the tagged packet at the tagged sensor consists of two parts: i) the remaining time of the current transmission cycle (with the packet arrival time as the beginning), and ii) the time spent by packets ahead of the tagged packet in the queue of the tagged sensor in the next transmission cycle. From the queueing model in Subsection \uppercase\expandafter{\romannumeral3}.B, as long as the Laplace-Stieltjes Transform (LST) and the second moment of the service time are derived, a closed-form expression of the queueing delay can be obtained. In the following, we first derive the approximate expression of the LST of the service time.

    The service time of packets in a sensor is i.i.d., whose distribution function is given by $B_s(t)$. Then, the LST of the service time can be calculated by $ B_s^*(z) =\int_{t=0}^{\infty}{e^{-zt}dB\left( t \right)}$. Since time is divided into equal-sized slots,
    the probability mass function of the service time (in timeslots) of the $i$-th packet in a sensor can be expressed as
    \begin{equation}
    	 f_{B_i}\left( k;\mathbf{P}_{\rm cov}^{\rm M} \right) = \mathbb{P}\left( B_i=k \right)
    	=\left( 1-\mathbf{P}_{\rm cov}^{\rm M} \right) ^{k-1}\mathbf{P}_{\rm cov}^{\rm M}, k = 1,2,\cdots,
    \end{equation}
    where $\mathbf{P}_{\rm cov}^{\rm M}$ is the coverage probability of the typical MDC as given in Theorem 1, and $k$ is the number of timeslots required to successfully transmit a packet to an MDC. Hence, the LST of the service time (in timeslots) can be approximated as
   \begin{equation}
      {\rm B}^*\left( z \right) \approx {\rm M}_{B_i}\left(-z\right) = \sum_{k=1}^{\infty}{\left( 1-\mathbf{P}_{\rm cov}^{\rm M} \right) ^{k-1}\mathbf{P}_{\rm cov}^{\rm M} e^{-zk}}
   	 =\frac{\mathbf{P}_{\rm cov}^{\rm M} e^{-z}}{1-\left( 1-\mathbf{P}_{\rm cov}^{\rm M} \right) e^{-z}},
    \end{equation}
 	 where ${\rm M}_{B_i}\left(z\right)$ is moment generating function of service time $B_i$. Thus, the second moment of the service time (in timeslots) of a packet is given by
    \begin{equation}\label{Eq_b2M}
      b^{(2)} = {\rm M}_{B_i}^{''}\left( 0 \right) = \frac{2 - \mathbf{P}_{\rm cov}^{\rm M}}{(\mathbf{P}_{\rm cov}^{\rm M})^2}.
    \end{equation}

    In the following, we derive the queueing delay at the tagged sensor in Theorem 3.

    \textbf{Theorem 3.} {\itshape When the system is in steady state, i.e. , $\frac{\xi}{\mu} < \mathbb{P}_{ct}$, the queueing delay of a packet at the tagged sensor is given by}\\
    \begin{equation}\label{FML_waitTime}
        \begin{aligned}
            {\rm E}\left( \mathbf{D}_q^s \right) =\left[\frac{\xi(2 - \mathbf{P}_{\rm cov}^{\rm M})}{2\mathbf{P}_{\rm cov}^{\rm M}(\mathbf{P}_{\rm cov}^{\rm M} - \xi)}+\frac{{\rm E}\left( L^* \right)}{\xi}\right]\delta,
        \end{aligned}
    \end{equation}
     {\itshape where $\mu = \mathbf{P}_{\rm cov}^{\rm M}$, and ${\rm E}\left( L^* \right)$ is given in Lemma 3.}

    {\itshape Proof}. According to Little's law, the average sojourn time of a packet in the tagged sensor is given by $ {\rm E}(T) = \frac{{\rm E}(L)}{\xi} = \mathbf{D}_q^s + \frac{1}{\mu}$, where the average queue length ${\rm E}(L^\ast)$ is given in Lemma 4. Thus, the average queueing length can be derived.

\subsubsection{Transmission Delay at the Tagged Sensor}

    The transmission delay at the tagged sensor is defined as the number of timeslots for a packet to be successfully received by an MDC, which is related to the coverage probability of the MDC. We assume that the packet transmission is independent in different timeslots.
    Thus, the number of required timeslots for a packet to be successfully transmitted follows a geometric distribution with $\mathbf{P}_{\rm cov}^{\rm M}$ as the success probability. The \textnormal{pmf} of number of slots required to deliver a packet is given by
    \begin{equation}\label{Eq_pmf}
        \mathbb{P}\left[T_k = m \right] = \mathbf{P}_{\rm cov}^{\rm M}\left(1-\mathbf{P}_{\rm cov}^{\rm M}\right)^{m-1}, \text{for}\ m = 1,2,\cdots.
    \end{equation}
    Thus, the mean transmission delay of a packet at the tagged sensor can be derived as
    \begin{equation}\label{Eq_E_Dts}
      {\rm E}(\mathbf{D}_t^s) = \frac{\delta}{\mathbf{P}_{\rm cov}^{\rm M}}.
    \end{equation}

\subsubsection{Queueing Delay at the Typical MDC}

    When the tagged packet is received by the typical MDC, it will be queued at the MDC and waits for being forwarded to an AP. According to the mobility model given in Section \uppercase\expandafter{\romannumeral2}.C, an MDC needs to collect a certain number of packets (denoted by $K$), before moving straightly to its associated AP, as shown in Fig. \ref{vacationModel}. During the data collection stage, the MDC moves according to the SRWP model, and encounters multiple sensors to collect packets. Thus, the queueing delay of the tagged packet at the typical MDC is mainly composed of the time interval for wandering and collecting packets from sensors, and the time interval for traveling to the aggregation area, where the former is mainly determined by number of collected packets $K$, the density of sensors, the packet arrival rate of the sensor $\xi$, the contact radius $R_s$ with a sensor, while the latter is related to the density of APs and the velocity of the MDC.

    The queueing delay caused by the packets collection from sensors is mainly determined by the average number of contacts $N_c$ (given in Eq. (\ref{eq_Nc})) between the typical MDC and sensors. Since the tagged packet is randomly selected, it may be collected by the typical MDC in any one of the $N_c$ contact durations with equal probability. We assume that the tagged packet is collected at the $i$-th contact duration, then the MDC still needs to collect the remaining packets for $N_c-i$ times on average, $i=1,2,\cdots,N_c$. The time interval between consecutive data collection periods is given by $T_{c} = {\rm E}(CT)+{\rm E}(ICT_s)$. For the contact during which the tagged packet is collected, denoted by $[t, t+{\rm E}(CT)]$, the timeslot $t_0$ at which the tagged packet is collected is assumed to be uniformly distributed within the contact duration, i.e., $t_0\sim {\rm U}[t,t+{\rm E}(CT)]$. Once $K$ packets are collected by the MDC, the MDC associates to the nearest AP, and moves straightly to the aggregation area of the AP. Hence, the queueing delay of the tagged packet at the typical MDC is derived as
    \begin{equation}\label{FML_mov_delay}
        \begin{aligned}
		  {\rm E}(\mathbf{D}_q^m)&= \frac{1}{N_c}(N_c-1 + N_c-2 + \cdots + 0)\cdot \left\{{\rm E}(CT) + {\rm E}(ICT_s)\right\} +\frac{{\rm E}(CT)}{2} + {\rm E}(T_{\rm Smov})\\
                           &= \frac{K\left\{{\rm E}(CT) + {\rm E}(ICT_s)\right\}}{\xi({\rm E}(CT) + {\rm E}(ICT))} - \frac{{\rm E}(ICT_s)}{2}
                           + \int_{R_a}^{\infty}{\frac{r}{v} 2\pi\lambda_b r \exp\left(-\pi \lambda_b r^2 \right) dr}.
	    \end{aligned}
    \end{equation}
where the first part denotes the queueing delay at the MDC caused by collection for $K$ packets, and the second part, i.e., ${\rm E}(T_{\rm Smov})$ denotes the queueing delay at the MDC caused by the straight movement.

\subsubsection{Transmission delay at the typical MDC}

    The transmission delay at the typical MDC is defined as the number of timeslots for a packet to be successfully received by an AP, which is related to the coverage probability of the AP. Thus, the mean transmission delay of a packet at the typical MDC can be derived as
    \begin{equation}\label{Eq_E_Dtm}
      {\rm E}(\mathbf{D}_t^m) = \frac{\delta}{\mathbf{P}_{\rm cov}^{\rm A}}.
    \end{equation}
    Altogether, the end-to-end delay is expressed as
     \begin{equation}\label{Eq_TotalDelay}
    	\begin{aligned}
    		{\rm E}(\mathbf{D}) &= {\rm E}(\mathbf{D}_q^s) + {\rm E}(\mathbf{D}_t^s) + {\rm E}(\mathbf{D}_q^m) + {\rm E}(\mathbf{D}_t^m) \\
    		&= \left[\frac{\xi(2 - \mathbf{P}_{\rm cov}^{\rm M})}{2\mathbf{P}_{\rm cov}^{\rm M}(\mathbf{P}_{\rm cov}^{\rm M} - \xi)}+\frac{{\rm E}\left( L^* \right)}{\xi}\right]\delta + \frac{\delta}{\mathbf{P}_{\rm cov}^{\rm M}}
    		+ \frac{K\left\{{\rm E}(CT) + {\rm E}(ICT_s)\right\}}{\xi({\rm E}(CT) + {\rm E}(ICT))} - \frac{{\rm E}(ICT_s)}{2}\\
    		&\quad+ \int_{R_a}^{\infty}{\frac{r}{v} 2\pi\lambda_b r \exp\left(-\pi \lambda_b r^2 \right) dr} + \frac{\delta}{\mathbf{P}_{\rm cov}^{\rm A}},
    	\end{aligned}
    \end{equation}
    where ${\rm E}\left( L^* \right)$ is given in Lemma 3.
\subsection{Energy Consumption}

    In this section, the energy consumption from both a sensor's perspective and a network's perspective is taken into consideration. From a sensor's perspective, due to the use of sleeping mode, the energy consumption consumed by a sensor (denoted by $\mathbf{E}_s$) is composed of two parts: (1) the energy consumed by a sensor to successfully transmit a packet to an MDC in the data collection stage, and (2) the energy consumed by a sensor in sleeping mode (normalized by the number of packets collected in a contact period). Therefore, we have
     \begin{equation}\label{Eq_Ener_s}
            \mathbf{E}_s = \frac{P_s\delta}{\mathbf{P}_{\rm cov}^{\rm M}} + \frac{{\rm E}(V)}{{\rm E}(\Psi)}P_{\rm sleep}
                         = \frac{P_s\delta}{\mathbf{P}_{\rm cov}^{\rm M}} + \frac{1-\rho}{\xi}P_{\rm sleep},
    \end{equation}
    where $\mathbf{P}_{\rm cov}^{\rm M}$ is the coverage probability of a typical MDC, ${\rm E}(V)$ represents the average vacation time, ${\rm E}(\Psi)$ represents the average number of packets collected by an MDC in a contact period, $P_s$ and $P_{\rm sleep}$ are power consumed in transmission mode and sleeping mode, respectively.

     From a network's perspective, we define the energy consumption as the average energy consumed by sensors and MDCs in a unit area when they successfully transmit a packet, denoted by $\mathbf{E}_n$. Similarly, the energy consumption from a network's perspective depends on the transmit power, the densities of active sensors and MDCs, and the transmission time at both the sensor and the MDC. Therefore, we have
    \begin{equation}\label{eq_ener_MDC_net}
      \mathbf{E}_n = \lambda_s^{'}(\frac{P_s\delta}{\mathbf{P}_{\rm cov}^{\rm M}} + \frac{1-\rho}{\xi}P_{\rm sleep}) + \frac{\lambda_m^{'} P_m\delta}{\mathbf{P}_{\rm cov}^{\rm A}},
    \end{equation}
    where $\mathbf{P}_{\rm cov}^{\rm A}$ is the coverage probability of a typical AP. $P_m$ represents the transmit power of the MDC, and $\lambda_s^{'}$ and $\lambda_m^{'}$ are the density of active sensors and MDCs, which are given in Eq. $(\ref{Eq_act_lamdas})$ and Eq. $(\ref{Eq_act_lamdam})$, respectively.
    		
\subsection{Validation}

    In this subsection, we perform extensive simulations over a square plane of $1000m \times 1000m$ to verify the accuracy of theoretical analysis results. 
    The transmit power of sensors and MDCs are set to $5 {\rm mW}$ and $10 {\rm mW}$ respectively. The thermal noise power is set to $-121\ {\rm dBm}$ and the path loss exponent is set to $3$. The radius of aggregation area is fixed to $20m$ and the contact radius of sensors is set to $10m$. The packets arrival rate is set to $0.6\ {\rm packets/s}$. The above simulation parameters are suitable for applications of agricultural IoT network, such as the farmland environment monitoring, monitoring for wildlife habitat, etc. In the following, the coverage performance analysis of the typical MDC and the typical AP is validated first. Then, we evaluate the queueing delay at the tagged sensor and the queueing delay at the typical MDC, respectively, by varying the velocity of MDCs.

\begin{figure}
    	\centering
    	\includegraphics[width=2.8in]{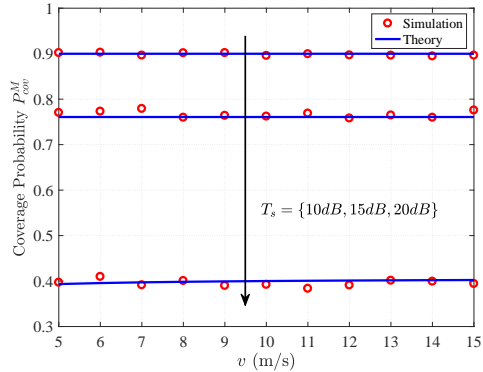} 
    	\caption{Coverage probability $P_{\rm cov}^{\rm M}$ vs. velocity of MDCs, for $\lambda_{s}= 10^{-3}m^{-2}$ and $\lambda_{m}= 5\times10^{-4} m^{-2}$} 
    	\label{figPrMdB}
    \end{figure}

    Figure \ref{figPrMdB} shows the coverage probability of the typical MDC as a function of velocity of MDCs for different SINR threshold. We find that simulation results match well with the result of theoretical analysis, which verifies accuracy of Theorem 1. We observe that the velocity of MDCs has little impact on the coverage probability $\mathbf{P}_{\rm  cov}^{\rm M}$. This can be explained by the fact that the increase of $v$ reduces the contact time and inter-contact time to the same extent, which makes the contact probability sensors almost unchanged. Moreover, given the packet arrival rate, when the system reaches a steady state, the simultaneous decrease of service time and vacation time will also keep the queue length nearly unchanged. Hence, the density of active sensors or aggregated interference from transmitting sensors in the network has little variation with the changing of velocity.

    \begin{figure}
        	\centering
        	\includegraphics[width=2.8in]{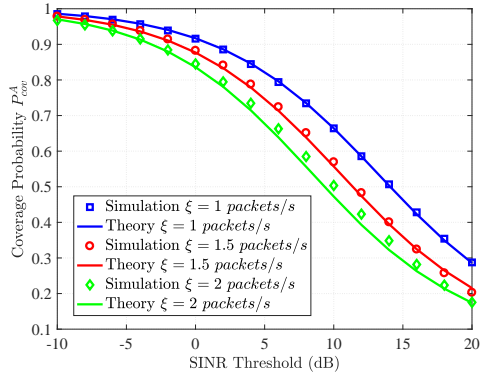} 
        	\caption{Coverage Probability $P_{\rm cov}^{\rm A}$ vs. SINR threshold, for $v = 5m/s$, $\lambda_{s} = 2 \times 10^{-3}m^{-2}$, $T_s = 0 {\rm dB}$, $\lambda_m = 1 \times 10^{-3}m^{-2}$, $K = 128\ {\rm packets}$, $\lambda_b = 4 \times 10^{-4}m^{-2}$, and $T_a = 0 {\rm dB}$ } 
        	\label{figPrCovBs}
    \end{figure}
    Figure \ref{figPrCovBs} depicts the coverage probability of the typical AP as a function of SINR threshold. The theoretical results are very close to the simulation results, which validates the accuracy of Theorem 2. We observe that a larger packet arrival rate $\xi$ leads to a smaller coverage probability of the typical AP. This comes from the fact that for larger values of packet arrival rate, MDCs can collect more packets from a sensor in a single contact duration, and come to a full buffer state (i.e., collecting $K$ packets) earlier, which leads to a higher active probability of MDCs, resulting in larger aggregated interference at the typical AP.

\begin{figure}[t]
    	\subfigure[]{
    		\begin{minipage}[t]{0.5\linewidth}
    			\centering
    			\includegraphics[width=2.7in]{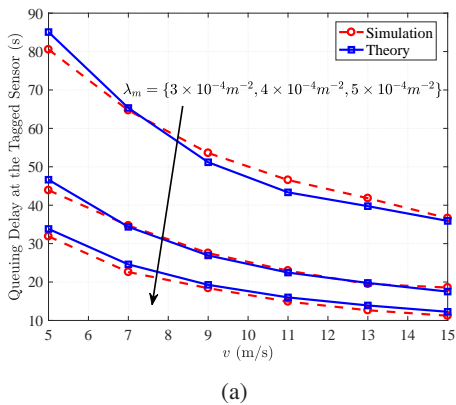}
                \label{figWaitingDelay}
    		\end{minipage}%
    	}
    	\subfigure[]{
    		\begin{minipage}[t]{0.5\linewidth}
    			\centering
    			\includegraphics[width = 2.7in]{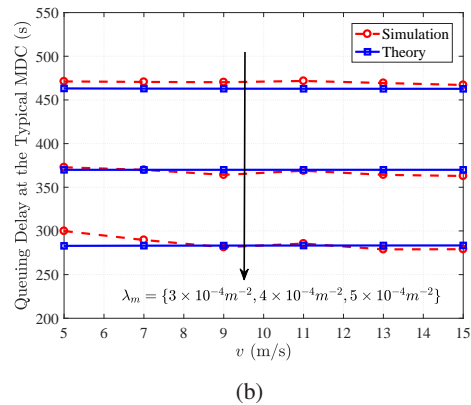}
                \label{figMovementDelay}
    		\end{minipage}%
    	}
    	\caption{Comparison of simulations and analytical results for (a) Queueing delay at the tagged sensor and (b) Queueing delay at the typical MDC v.s. MDC velocity $v$, for $\lambda_s = 5 \times 10^{-4}m^{-2}$, $T_s = 10 {\rm dB}$, $T_a = 0 {\rm dB}$, $\lambda_b = 1 \times 10^{-4}m^{-2}$,  $K = 512\ {\rm packets}$.}
\end{figure}		
		
    Figure \ref{figWaitingDelay} shows the queueing delay at the tagged sensor as a function of velocity of MDCs for different densities of MDCs $\lambda_m$. A good match between the numerical results and the simulation results confirms the accuracy of Theorem 3. we also observe that with the increase in velocity $v$ or density of MDCs $\lambda_m$ , the queueing delay at the tagged sensor decreases. The change is chiefly caused by the fact that a high-velocity moving MDC or high-density MDCs can provide more contact opportunities with sensors, reducing the queueing delay of packets at the tagged sensor. However, when the velocity of MDCs achieves a certain value (10 m/s in this example), the change in queueing delay at the tagged sensor is very small. This results from the fact that given contact radius of sensors, a higher velocity of the MDC leads to a smaller contact period, which increases the number of contact with sensors for collecting $K$ packets. Therefore, when the velocity of MDCs exceeds a certain value, the queueing delay at the tagged sensor remains almost unchanged.

    Figure \ref{figMovementDelay} depicts the queueing delay of the tagged packet at the typical MDC as a function of velocity of MDCs for different densities of MDCs $\lambda_m$. Small deviation between theoretical results and simulation results verifies the accuracy of Eq. (\ref{FML_mov_delay}). It reveals the fact that the velocity of MDCs has little effect on the queueing delay of the tagged packet at the typical MDC. It can be explained by the balanced opposite effects of MDC's velocity on the average inter-contact duration ${\rm E}(ICT_s)$ and the number of packets $N_c$ collected within one contact duration. An increase in the velocity of the MDC, decreases ${\rm E}(ICT_s)$ between the MDC and sensors on one hand, and enlarges the required number of contact times $N_c$ to collect $K$ packets on the other hand. The opposite effects balance out each other, resulting in an unchanged queueing delay at the typical MDC.

\section{NUMERICAL RESULTS AND DISCUSSIONS}

    In this section, we analyze the impact of key system parameters on network performance, and then obtain the parameter value range that maximizes the system performance gain.

    \subsection{Impact of Density of Sensors}

\begin{figure}[t]
    	\subfigure[]{
    		\begin{minipage}[t]{0.5\linewidth}
    			\centering
    			\includegraphics[width=2.7in]{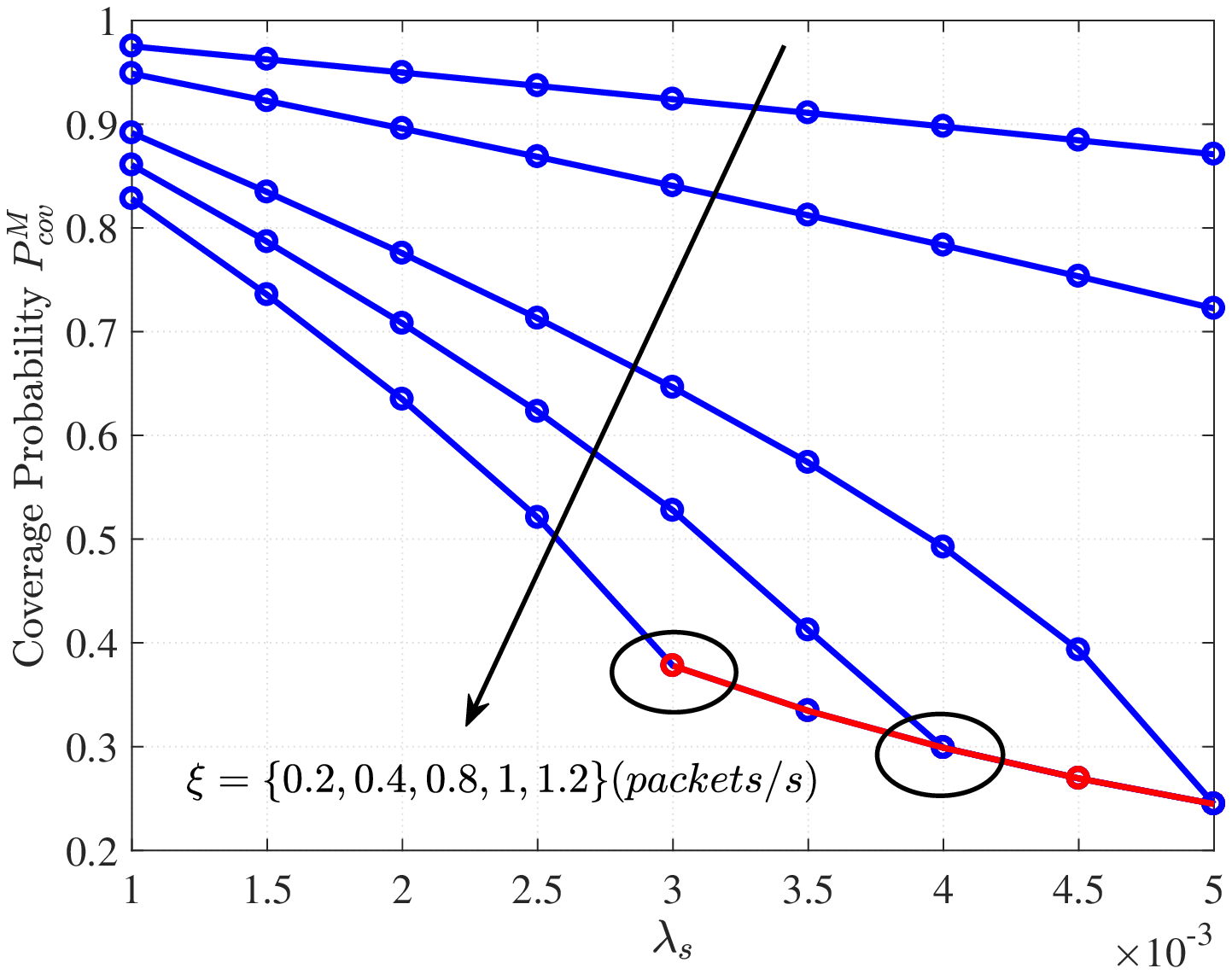}
                \label{figPrMlamdas}
    		\end{minipage}%
    	}
    	\subfigure[]{
    		\begin{minipage}[t]{0.5\linewidth}
    			\centering
    			\includegraphics[width =2.7in]{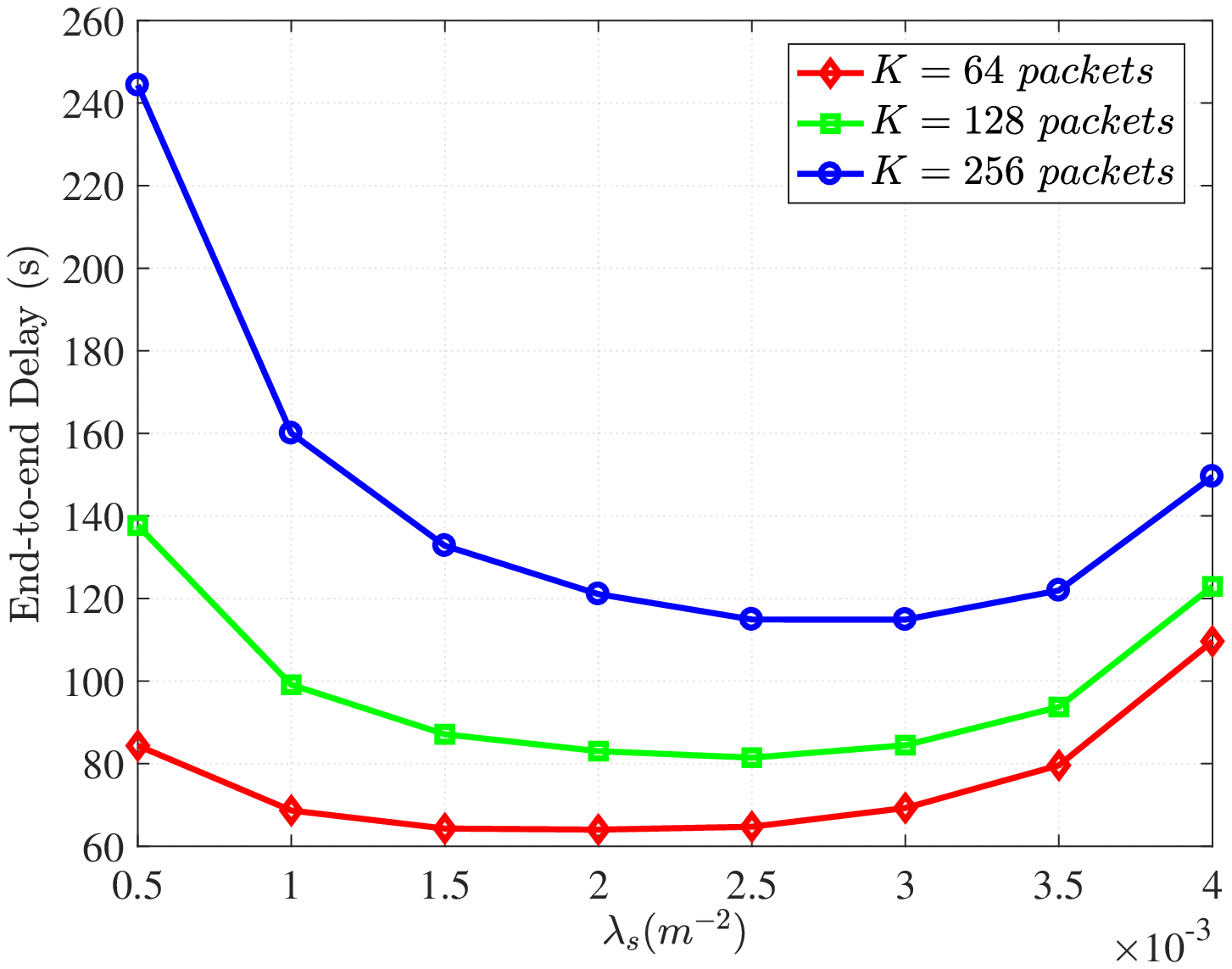}
                \label{figDelaylamdas}
    		\end{minipage}%
    	}
    	\caption{Coverage probability  $P_{\rm cov}^{\rm M}$ and end-to-end delay vs. density of sensors, for $v = 5m/s, R_s = 10m$, $T_s = 10{\rm dB}$, $T_a = 0{\rm dB}$, $\lambda_{m} = 10^{-3}m^{-2}$ and $\lambda_{b} = 10^{-4}m^{-2}$.}
    \end{figure}

     Figure \ref{figPrMlamdas} depicts the coverage probability of the typical MDC $\mathbf{P}_{\rm cov}^{\rm M}$ as a function of density of sensors $\lambda_s$ for different packet arrival rate $\xi$. We find that the coverage probability shows a steady decline with the increase of $\lambda_s$. This is due to the fact that, as the density of sensors grows, the aggregated interference from transmitting sensors in the network goes up, leading to a decrease of the coverage probability or the service rate. In addition, we observe that as the packet arrival rate increases, the coverage probability or the service rate decreases. This stems from the fact that as the packet arrival rate increases, the probability that the sensor queue is not empty goes up. It is worth noting that when the packet arrival rate is relatively high, the sensor falls into a fully loaded state when its density increases to a critical value which is marked by a circle in the figure. Hence, when the sensor density exceeds a critical value, the packet arrival rate will be greater than the service rate, and the system will be in an unstable state.

     The impact of sensor density on coverage probability $\mathbf{P}_{\rm cov}^{\rm M}$ further affects the end-to-end delay. In Fig. \ref{figDelaylamdas}, we depict the end-to-end delay as a function of the sensor density $\lambda_s$ for different MDC collection thresholds $K$. We can observe that as the sensor density increases, the end-to-end delay first decreases and then increases. This is because that the larger the $K$, the higher the queueing delay at the typical MDC will be. Meanwhile, in a network with a higher sensor density, MDCs take less time to collect a certain number of packets, reducing the queueing delay at the typical MDC. However, when the sensor density continues to increase and exceeds a certain value, the decreased coverage probability caused by the higher aggregated interference significantly enlarges the queueing delay at the tagged sensor, leading to the increasing end-to-end delay.

     \subsection{Impact of Contact Radius of Sensors}

\begin{figure}[t]
    	\subfigure[]{
    		\begin{minipage}[t]{0.5\linewidth}
    			\centering
    			\includegraphics[width=2.7in]{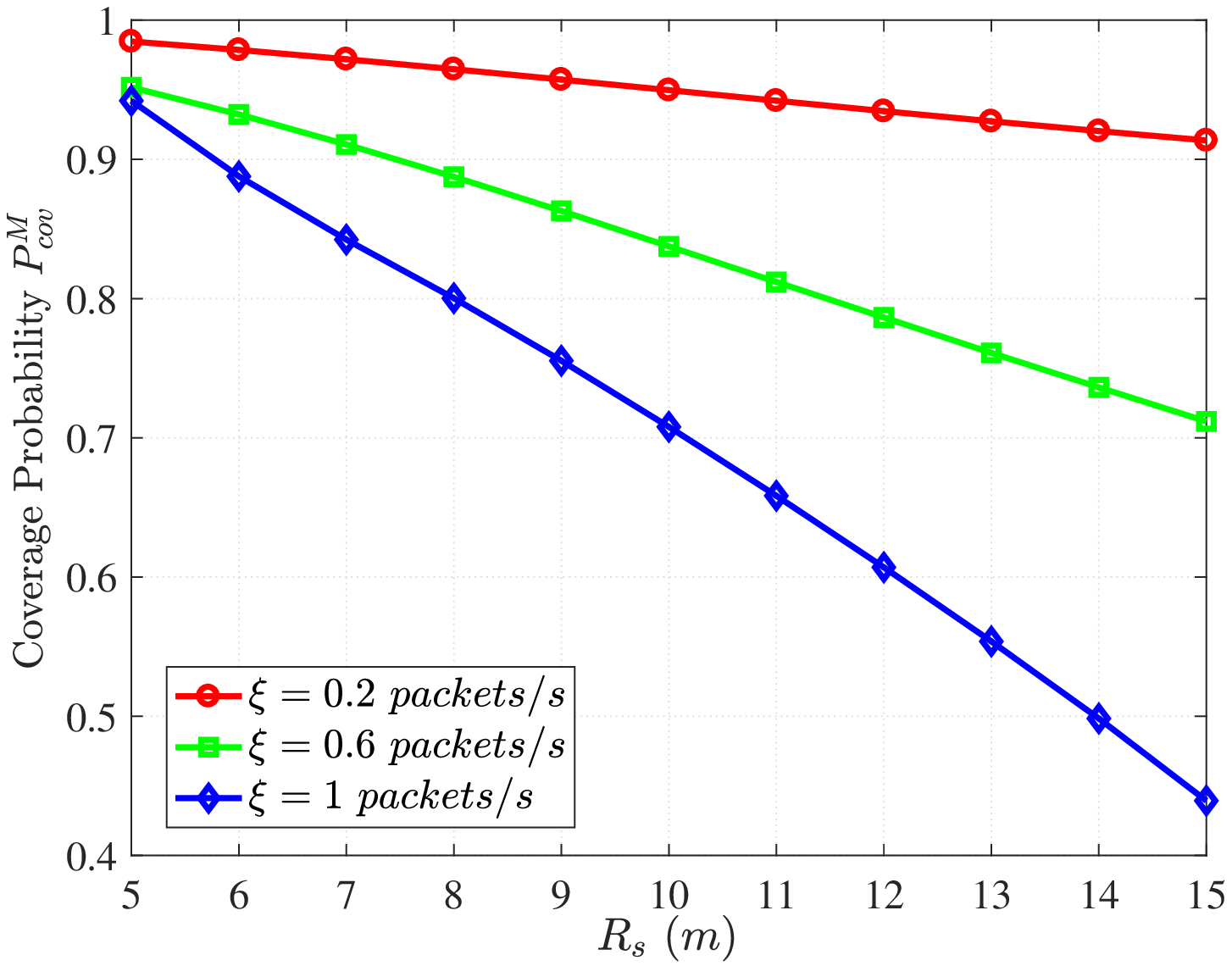}
                \label{figPrMRs}
    		\end{minipage}%
    	}
    	\subfigure[]{
    		\begin{minipage}[t]{0.5\linewidth}
    			\centering
    			\includegraphics[width =2.7in]{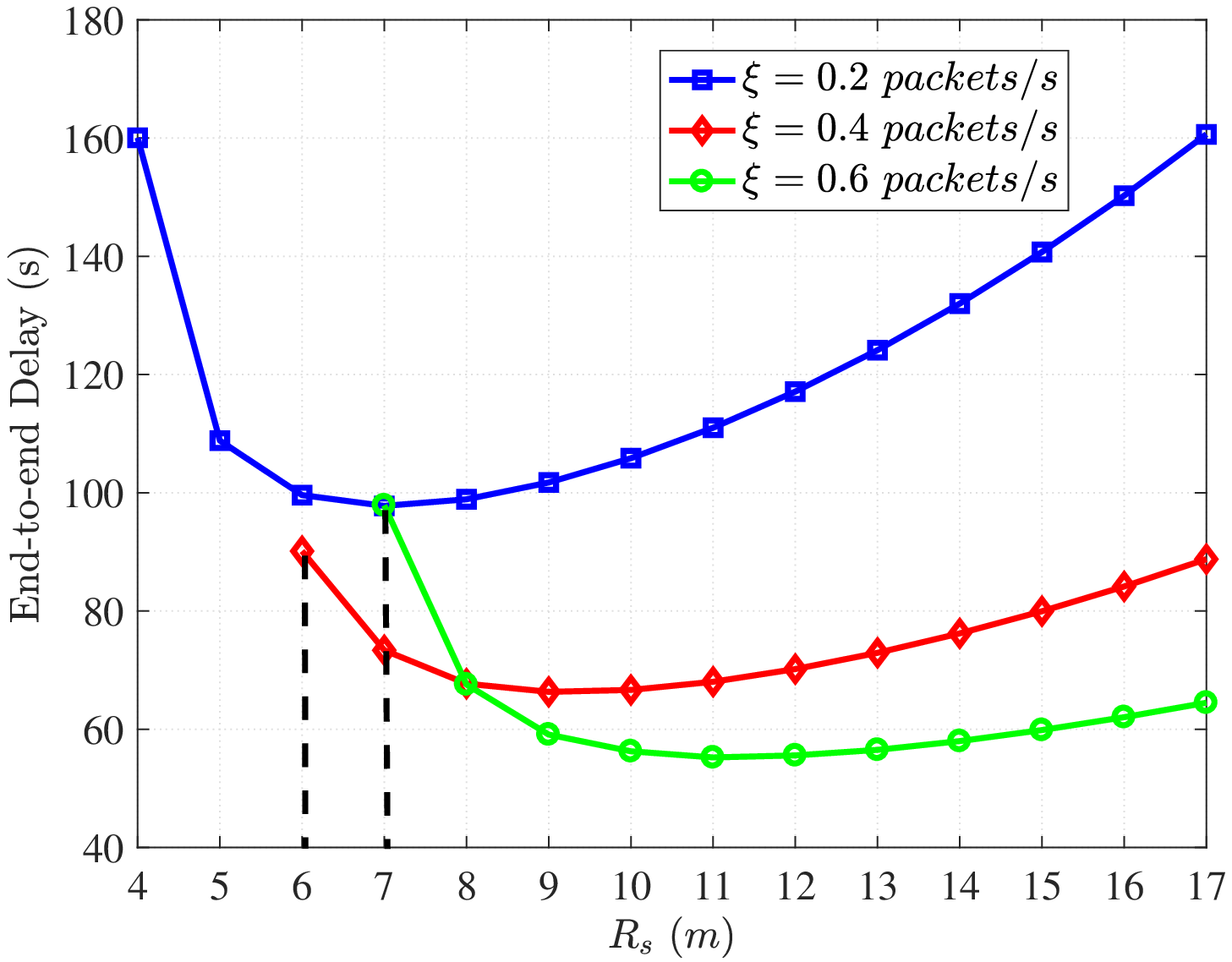}
                \label{TotalDelayRs}
    		\end{minipage}%
    	}
    	\caption{Coverage probability $P_{\rm cov}^{\rm M}$  and end-to-end delay vs. contact radius of sensors, for $v = 5m/s$, $\lambda_{s} = 2\times10^{-3}m^{-2}$, $T_s = 10{\rm dB}$, $\lambda_{b} = 10^{-4}m^{-2}$ and $K = 64\ {\rm packets}$, where $\lambda_{m} = 6\times10^{-4}m^{-2}$ in (a) and $\lambda_{m} = 10^{-3}m^{-2}$ in (b).}
    \end{figure}

     In Fig. \ref{figPrMRs}, we depict the coverage probability of the typical MDC as a function of the contact radius $R_s$ for different packet arrival rate $\xi$. It reveals the fact which the coverage probability declines with increase of contact radius, which results from the decreasing signal power and the growing aggregated interference. On the one hand, a larger contact radius leads to a decrease of signal power. On the other hand, as $R_s$ enlarges, ${\rm E}(ICT)$ decreases while ${\rm E}(CT)$ increases, which enlarges the density of active sensors and leads to the increase in aggregated interference. We can also observe that, for a given $R_s$, the coverage probability of the typical MDC decreases with the increasing packet arrival rate. It follows the fact that, when the packet arrival rate increases, the probability that the queue of sensors is non-empty increases and the active density of sensors increases, leading to the growing aggregated interference. The impact of contact radius $R_s$ on coverage probability $\mathbf{P}_{\rm cov}^{\rm M}$ further affects the total delay.

     The Figure \ref{TotalDelayRs} shows the total delay as a function of $R_s$ for different packet arrival rates $\xi$. The dotted line in the figure indicates the minimum $R_s$ that keeps the system in a steady state under the corresponding packet arrival rate. We can observe that, for a given packet arrival rate, the total delay first declines and then rise up with the increase of $R_s$. It can be explained by the fact that as $R_s$ enlarges, ${\rm E}(CT)$ increases, as a result, both the queueing delay at the tagged sensor and the queueing delay at the typical MDC decreases. However, when the $R_s$ is higher than a certain value, the coverage probability $\mathbf{P}_{\rm cov}^{\rm M}$ or equivalently the service rate decreases, leading to the increase of the end-to-end delay.

\subsection{Impact of Density of MDCs and Velocity of MDCs}

\begin{figure}[t]
    	\subfigure[]{
    		\begin{minipage}[t]{0.5\linewidth}
    			\centering
    			\includegraphics[width=2.7in]{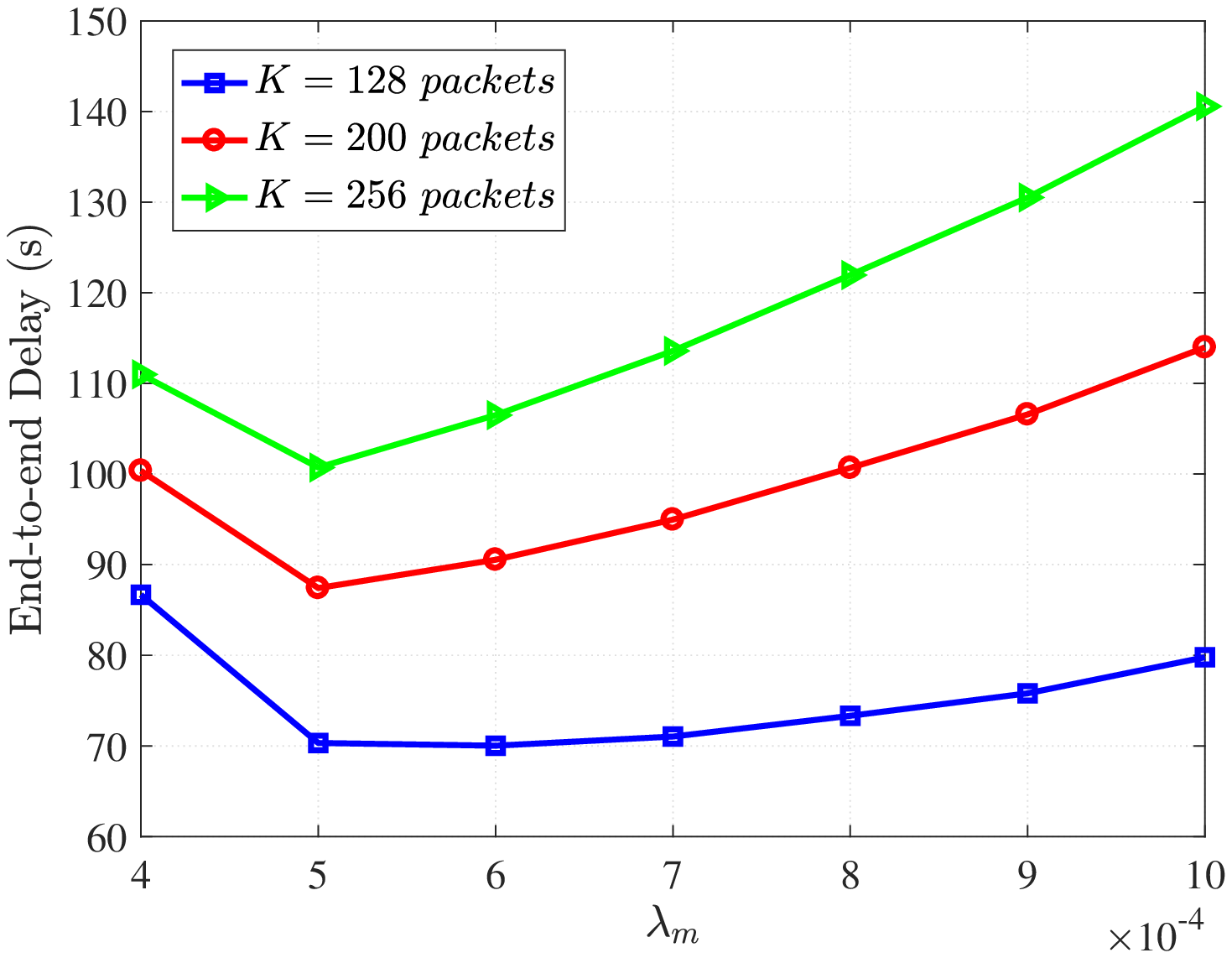}
                \label{figTotalDelay_lamdam}
    		\end{minipage}%
    	}
    	\subfigure[]{
    		\begin{minipage}[t]{0.5\linewidth}
    			\centering
    			\includegraphics[width =2.7in]{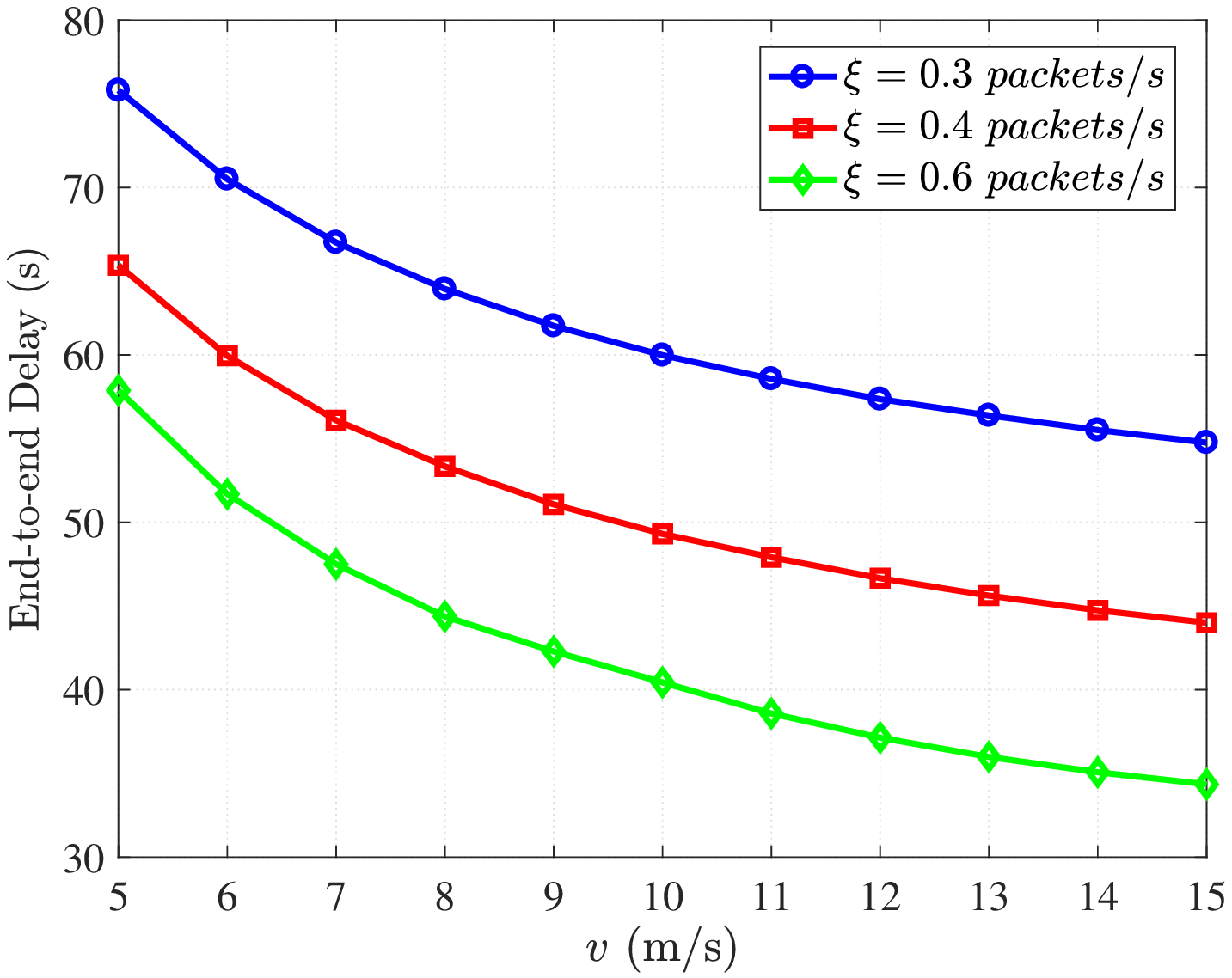}
                \label{figTotalDelay_v}
    		\end{minipage}%
    	}
    	\caption{End-to-end Delay as a function of MDCs' density (a) and MDCs' velocity (b), for $\lambda_s = 2 \times 10^{-3}m^{-2}$, $R_s = 10m$, $v = 5m/s$, $T_s = 10 {\rm dB}$, $T_a = 0{\rm dB}$ and $\lambda_a = 1\times 10^{-4}m^{-2}$, where $\xi = 0.6\ {\rm packets}/s$  in (a) and $K = 64\ \rm {packets}$ in (b).}
    \end{figure}

    Figure \ref{figTotalDelay_lamdam} depicts the end-to-end delay as a function of density of MDCs under different MDC collection threshold $K$. It can be seen that as the MDC density increases, the end-to-end delay begins to decrease rapidly. When the MDC density reaches a certain value, the end-to-end delay gradually increases. This comes from the fact that, the end-to-end delay is mainly determined by the queueing delay of at the tagged sensor and the queueing delay at the typical MDC. As the MDC density increases, it is obvious that the queueing delay of at the tagged sensor decreases. But when the density of MDC increases to exceed a certain threshold, the movement delay of a packet will increase, This is due to the fact that when the network is in steady state, a higher MDC density magnifies the competition of data collection among MDCs, which decreases the number of packets can be collected in each contact period. To meet the requirement of collecting $K$ data packets, more sensors must be contacted, enlarging the queueing delay at the MDC. Therefore, Fig. \ref{figDelaylamdas}, \ref{TotalDelayRs}, \ref{figTotalDelay_lamdam} depict that the proposed analytical framework can be applied to minimize the end-to-end delay by optimizing the density of sensors, MDCs, and the contact radius $R_s$.

    Figure \ref{figTotalDelay_v} depicts the end-to-end delay as a function of MDCs' velocity under different packet arrival rate. It can be seen that with the increase of MDCs' velocity, the end-to-end delay decreases. This comes from the fact that, as the velocity of MDCs increases, the average queueing delay at the tagged sensor, i.e., $\mathbf{D}_q^s$ decreases, and the queueing delay at the typical MDC, i.e., $\mathbf{D}_q^m$ is nearly unchange. Hence, the end-to-end delay decreases.

\subsection{Energy Consumption Comparison}

%

\begin{figure}[t]
        \subfigure[]{
    		\begin{minipage}[t]{0.5\linewidth}\label{figEnerSensor}
    			\centering
    			\includegraphics[width = 2.7in]{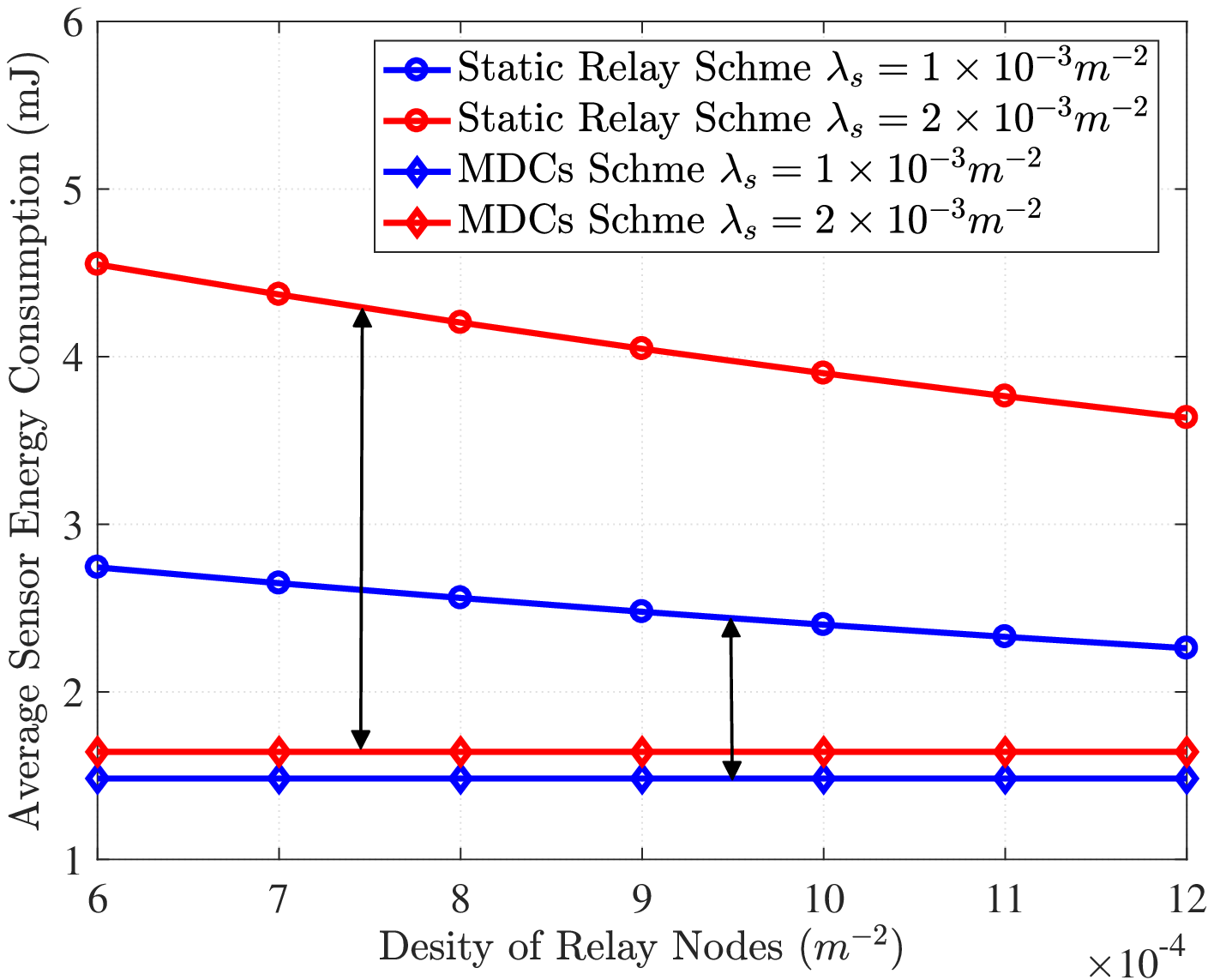}
    		\end{minipage}%
    	}
        \subfigure[]{
    		\begin{minipage}[t]{0.5\linewidth}\label{figEnerNet}
    			\centering
    			\includegraphics[width = 2.7in]{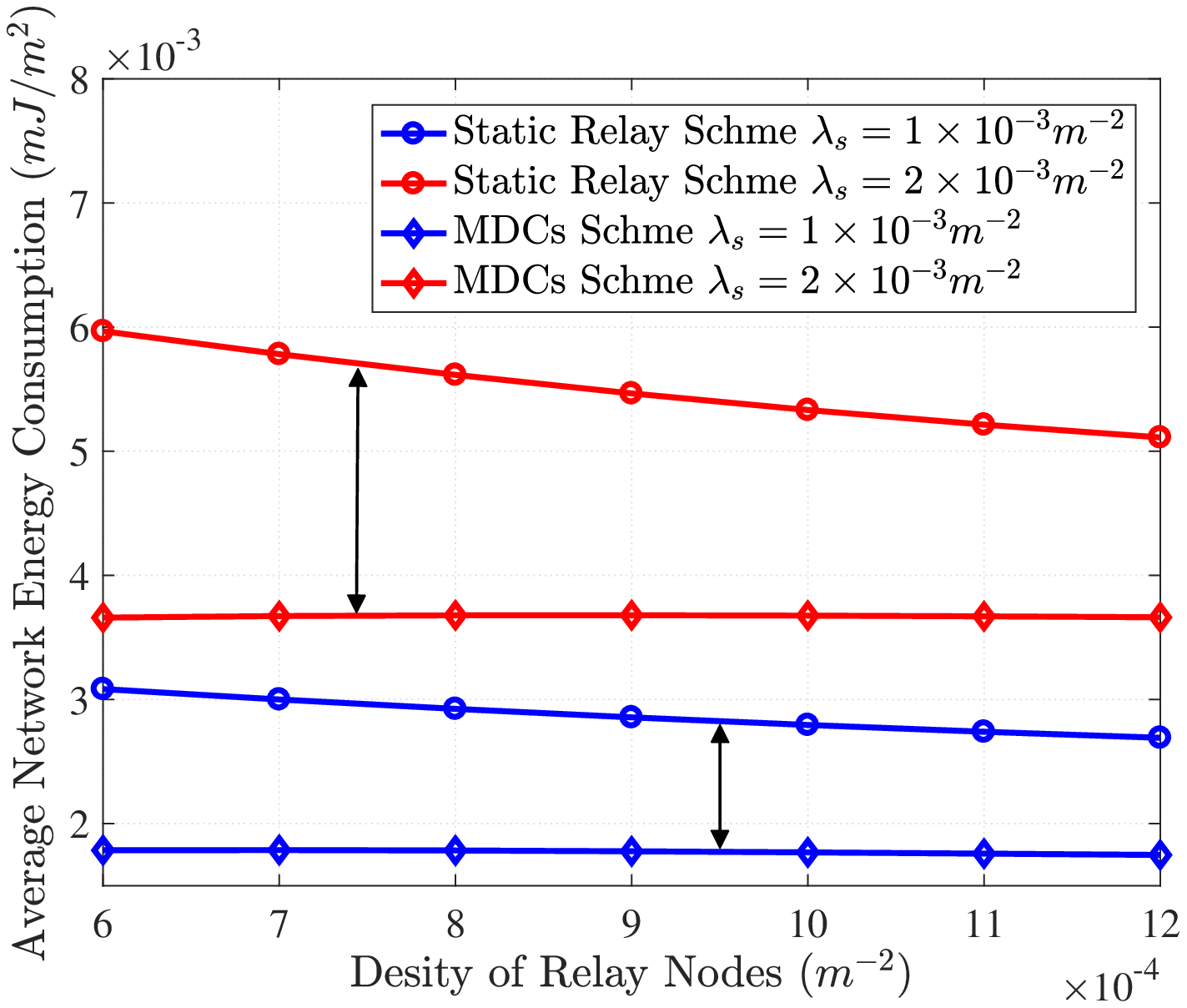}
    		\end{minipage}%
    	}
    	\caption{Energy Consumption vs. density of MDCs or static relays, for $\xi = 1\ {\rm packets/s}, v = 5m/s,R_s = 10m$, and $\lambda_b = 4 \times 10^{-4}m^{-2}$}
        \label{figEner}
    \end{figure}

     In Fig. \ref{figEner}, we depict the energy consumption as a function of density of relay nodes for different sensor density. To show the superiority of our proposed MDC scheme, we compare with the scheme proposed in \cite{kim2017delay} where static relay nodes are deployed to forward the packet from sensors to the base station.

    Figure \ref{figEnerSensor} depicts the average sensor energy consumption of delivering a packet as a function of relay density under the static relay scheme proposed in \cite{kim2017delay} and our proposed MDC scheme. It reveals the fact that with the help of randomly moving MDCs, the energy consumption of the sensor can be effectively reduced. In addition, with the increase of sensor density, the energy consumption of sensors in the static relay network increases significantly, while the energy consumption of sensors with the proposed MDC scheme does not change a lot. This is because that for the static relay scheme, network collisions increase sharply with the density of active sensors, which reduces the successful transmission probability. Moreover, the static relay scheme is highly dependent on the density of the relay. The higher the static relay density, the greater the probability of successful transmission probability and the less energy the sensor consumes. To the contrary, the proposed MDC scheme greatly reduces the required number of relays, which also decreases the aggregated interference in the data aggregation stage. Figure \ref{figEnerNet} depicts average network energy consumption as a function of relay density under the static relay scheme and our proposed MDC scheme, which shows the similar trend as that in Fig. \ref{figEnerSensor}, and can be explained following the similar line as that for Fig. \ref{figEnerSensor}.

\section{Conclusion}
    In this study, we focused on an MDC-enhanced IoT network and proposed a theoretical framework to analyze the network performance in terms of coverage probability, end-to-end delay, and energy consumption. We adopted the SRWP mobility model for the MDC in the data collection stage, and modeled the data collection system between a sensor and MDCs as an M/G/1 vacation queueing system general limited (G-limited) service. By quantifying the effect of key parameters on the network performance, we concluded that the velocity of MDCs has little impact on coverage probability, while the end-to-end delay can be minimized by optimally setting the density and contact radius of sensors, and the velocity and density of MDCs. When the network is in a steady state, there is an upper bound on the arrival rate of the sensor packets, which is related to the contact probability and coverage probability of the MDC. To make the network keep stable, a higher sensor packet arrival rate requires a higher receiver sensitivity so as to enlarge the sensor contact area and thus the average contact time between the sensor and MDCs. There are several interesting directions for future work. One possible direction would be to consider the prioritized transmissions by considering packets of different priorities. Another possible direction would be to incorporate the heterogeneity of MDCs in terms of storage, computing capability and moving speed, and reveal the impact of such heterogeneity on the network performance.
    	
	
 \begin{appendices}

    \section{PROOF OF LEMMA 1}

Since all MDCs follow the SRWP mobility model and are independent of each other, for a typical MDC, when it passes through the contact area of the sensor which is defined as a circle with radius $R_s$, there are two events that occur mutually exclusive. One of event is that the MDC crosses in a straight line without sojourn, which is denoted by ${\rm W}$, the other event is that the MDC sojourns at a random position of contact area for a fixed time $p$, and then select a random angle $\theta \in [0,2\pi]$ to leave the contact area, which is denoted by ${\rm P}$. The duration of events ${\rm W}$ and ${\rm P}$ are denoted by $T_w$ and $T_p$, respectively. According to the total probability theorem, the average contact time can be obtained as
\begin{equation}\label{proof_E_CT}
{\rm E}(CT) = (1-\mathbb{P}_p)\cdot {\rm E}(T_w)+\mathbb{P}_p\cdot {\rm E}(T_{p}),
\end{equation}
where $\mathbb{P}_p$ denotes the probability of the MDC sojourns in the contact area of the sensor. The $\mathbb{P}_p$ can be given by
\begin{equation}\label{proof_Pp}
\mathbb{P}_p = \frac{\pi R_s}{2wv}, \qquad w > \frac{2R_s}{v},
\end{equation}
where $w$ represents the walk duration of the MDC as defined in Definition 2.
Due to the PDF of $T_w$ can be derived by
\begin{equation}
f_{T_{w}}(t)=\frac{v^{2} t}{2 R_s \sqrt{4 R_s^{2}-v^{2} t^{2}}}, \quad 0< t < \frac{2 R_s}{v},
\end{equation}
the expectation of $T_w$ can be obtained as ${\rm E}(T_w) = \frac{\pi R_s}{2v}$.
 \begin{figure}
	\centering
	\includegraphics[width=2.6in]{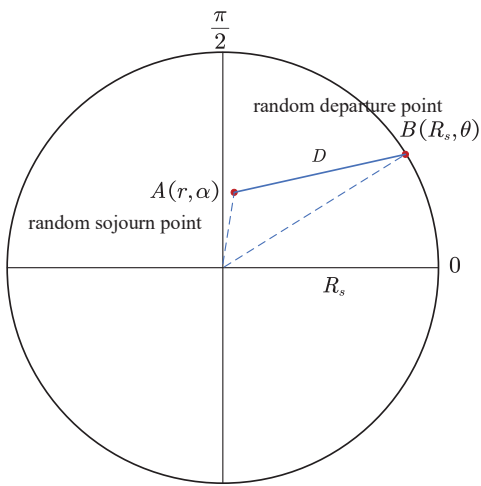} 
	\caption{Illustration of Angel in Lemma1} 
	\label{CircleAngel}
\end{figure}

When the event ${\rm P}$ occurs, the random sojourn position is denoted by ${\rm A}(r,\alpha)$ in the polar coordinate system. When the MDC leaves the circle, the intersection point between its trajectory and the circle is denoted by ${\rm B}(R_s,\theta)$, as shown in Fig. \ref{CircleAngel}. In addition, the distance between ${\rm A}$ and ${\rm B}$ is denoted by $D$. Since the coordinates of point ${\rm A}$ and point ${\rm B}$ follow the uniform distribution in the circle and on the circumference respectively, the expectation of $T_p$ can be derived by
 \begin{equation}\label{proof_Tp}
 {\rm E}(T_p) = \frac{2{\rm E}(D)}{v}+p,
 \end{equation}
where $v$ is the velocity of the MDC. The joint PDF of $(r,\alpha)$ is given by
\begin{equation}	
f_{\rm A}(r, \alpha)=\left\{\begin{array}{ll}\frac{r}{\pi R_{s}^{2}}, & 0 \leq r \leq R_{s}, 0 \leq \alpha \leq 2 \pi \\
	0, & \text { otherwise }
	\end{array}\right..
\end{equation}
In addition, the PDF of $\theta$ is given by $f_{\rm B}(\theta) = \frac{1}{2 \pi}, 0 \leq \theta \leq 2 \pi$. Due to $(r,\alpha)$ is independent of $\theta$, hence, the joint PDF of $(r,\alpha,\theta)$ is derived by
\begin{equation}
    f_{\rm A, B}(r, \alpha, \theta)=\left\{\begin{array}{l}\frac{r}{2 \pi^{2} R_{s}^{2}}, 0 \leq r \leq R_{s}, 0 \leq \alpha,\theta \leq 2 \pi\\
	0, \quad \text { otherwise }
	\end{array}\right..
\end{equation}
	Hence, the expectation of $D$ is derived by
\begin{equation}\label{proof_ED}
  {\rm E}\left(D\right)=\frac{1}{2 \pi^{2} R_{s}^{2}} \int_{0}^{2 \pi} \int_{0}^{2 \pi} \int_{0}^{R_{s}}
  \sqrt{\left(r \cos \alpha-R_{s} \cos \theta\right)^{2}+\left(r \sin \alpha-R_{s} \sin \theta\right)^{2}} r dr d\alpha d\theta.
\end{equation}
Substituting (\ref{proof_Pp}), (\ref{proof_Tp}) and (\ref{proof_ED}) into (\ref{proof_E_CT}), the proof is complete.

\section{PROOF OF THEOREM 1}
We assume that the typical MDC is located at a distance $r_0$ away from the tagged sensor ($r_0 \leq R_s$), according to Eq. (\ref{FML_SINR_define2}), the coverage probability of the typical MDC is derived by
$$
\begin{aligned}
&P_{\rm cov}^{\rm M}\left(T_s, \lambda_{s}, \lambda_{b}, R_s\right) ={\rm E}_{r_{0}}\left[\operatorname{\mathbb{P}}\left(SINR_m>T_s | r_{0}\right)\right]  \\
&=\int_{0}^{R_s} \operatorname{\mathbb{P}}\left[\frac{P_s h r_{0}^{-\alpha}}{I_r^m+\sigma^{2}} > T_s | r_{0}, I_r^m\right] \frac{2 r_{0}}{R_{s}^{2}} d r_{0}  \\
&=\int_{0}^{R_s} \mathbb{P}\left[h > P_s^{-1} T_s r_{0}^{\alpha}\left(I_r+\sigma^{2}\right) | r_{0}, I_r\right] \frac{2 r_{0}}{R_{s}^{2}} d r_{0},
\end{aligned}
$$
where $f_r(r_0)$ denotes the link distance between the typical MDC and the tagged sensor, which is given in (\ref{FML_fr0}). Due to the assumption of Rayleigh fading channel, i.e., $h \sim exp(1)$, the distribution in the above formula can be expressed as
$$\begin{aligned}
&\mathbb{P}\left[ h > P_s^{-1} T_s r_{0}^{\alpha}\left(I_r^m+\sigma^{2}\right) | r_{0}, I_r^m\right] \\
&={\rm E}_{I_r^m}\left[ h >P_s^{-1} T_s r_{0}^{\alpha}\left(I_r^m+\sigma^{2}\right) | r_{0}, I_r^m\right]  \\
&=\left.e^{-P_s^{-1} T_s r_{0}^{\alpha} \sigma^{2}} \cdot \mathcal{L}_{I_r^m}(s)\right|_{s = P_s^{-1} T_s r_{0}^{\alpha}},
\end{aligned}
$$
where $\mathcal{L}_{I_r^m}(s)$ is the LST of aggregated interference $I_r^m$. According to the definition of LST, we can get
$$
    \begin{aligned}
        &\mathcal{L}_{I_r^m}(s) = {\rm E}_{I_r^m}\left[e^{-s I_r^m}\right] \\
        &={\rm E}\left[\exp\left(-s\left(\sum_{x \in \Phi_{s}\backslash\left\{s_{0}\right\}} \mathbbm{1}_x  P_{s} h_{x} r_{x}^{-\alpha}\right)\right)\right]
    \end{aligned}
    $$
    $$
    \begin{aligned}
        &\mathop{=}\limits^{(a)}{\rm E}_{\Phi_{s}}\left[\prod_{x\in\Phi_{s}\backslash\left\{s_{0}\right\}} \left(\frac{\mathbb{P}_{act}^s}{1+s P_{s} r_{x}^{-\alpha}} + 1 - \mathbb{P}_{act}^s\right) \right] \\
        &\mathop{=}\limits^{(b)} \exp \left(-2 \pi \mathbb{P}_{act}^s\lambda_{s}\int_{r_{0}}^{\infty}\frac{T_s r_{x}}{T_s +\left(\frac{r_{x}}{r_{0}}\right)^{\alpha}} d r_{x}\right)
    \end{aligned}
$$
where step (a) is obtained according to the moment generation function (MGF) of $h_x$, and step (b) follows from the probability generating functional (PGFL) of a PPP.

\end{appendices}
\bibliography{reference}
\end{document}